\documentclass[twocolumn,aps,superscriptaddress,floatfix,prl,showpacs,nolongbibliography]{revtex4-2}
\usepackage{graphicx}
\usepackage{amsmath}
\usepackage{amssymb}
\usepackage{hyperref}
\usepackage{color}
\usepackage{braket}
\usepackage{dsfont}
\usepackage{lineno}
\usepackage{orcidlink}

\usepackage{float}

\newcommand{\Tr}{\operatorname{Tr}}

\begin{document}
\title{Ballistic‑to‑Localized Dynamics as Signature of Quantum Phase Transition in Josephson Junction}

\author{F. G. Capone\orcidlink{0009-0000-9832-4385}}
\affiliation{Dip. di Fisica E. Pancini - Università di Napoli Federico II - I-80126 Napoli, Italy}
\affiliation{INFN, Sezione di Napoli - Complesso Universitario di Monte S. Angelo - I-80126 Napoli, Italy}

\author{A. de Candia\orcidlink{0000-0002-9869-1297}}
\affiliation{INFN, Sezione di Napoli - Complesso Universitario di Monte S. Angelo - I-80126 Napoli, Italy}
\affiliation{SPIN-CNR and Dip. di Fisica E. Pancini - Università di Napoli Federico II - I-80126 Napoli, Italy}

\author{G. Di Bello\orcidlink{0000-0001-9838-6631}}
\affiliation{Dip. di Fisica E. Pancini - Università di Napoli Federico II - I-80126 Napoli, Italy}
\affiliation{INFN, Sezione di Napoli - Complesso Universitario di Monte S. Angelo - I-80126 Napoli, Italy}

\author{V.~Cataudella\orcidlink{0000-0002-1835-1429}}
\affiliation{INFN, Sezione di Napoli - Complesso Universitario di Monte S. Angelo - I-80126 Napoli, Italy}
\affiliation{SPIN-CNR and Dip. di Fisica E. Pancini - Università di Napoli Federico II - I-80126 Napoli, Italy}

\author{R. Fazio\orcidlink{0000-0002-7793-179X}}
\affiliation{Dip. di Fisica E. Pancini - Università di Napoli Federico II - I-80126 Napoli, Italy}
\affiliation{The Abdus Salam International Center for Theoretical Physics (ICTP), Strada Costiera 11, 34151 Trieste, Italy}

\author{N. Nagaosa\orcidlink{0000-0001-7924-6000}}
\affiliation{RIKEN Center for Emergent Matter Science (CEMS), Wako, Saitama 351-0198, Japan}
\affiliation{Fundamental Quantum Science Program (FQSP), TRIP Headquarters, RIKEN, Wako 351-0198, Japan}

\author{C. A. Perroni\orcidlink{0000-0002-3316-6782}}
\affiliation{INFN, Sezione di Napoli - Complesso Universitario di Monte S. Angelo - I-80126 Napoli, Italy}
\affiliation{SPIN-CNR and Dip. di Fisica E. Pancini - Università di Napoli Federico II - I-80126 Napoli, Italy}

\author{G. De Filippis\orcidlink{0000-0003-0557-3556}}
\affiliation{INFN, Sezione di Napoli - Complesso Universitario di Monte S. Angelo - I-80126 Napoli, Italy}
\affiliation{SPIN-CNR and Dip. di Fisica E. Pancini - Università di Napoli Federico II - I-80126 Napoli, Italy}

\begin{abstract}
Using state-of-the-art numerical techniques, we investigate how quantum phase fluctuations and quasiparticle tunneling shape the behavior of a small-capacitance Josephson junction across Ohmic, sub-Ohmic, and super-Ohmic dissipation regimes. We show that increasing the Ohmic dissipation strength drives a Berezinskii–Kosterlitz–Thouless quantum phase transition at thermodynamic equilibrium. Deviations from Ohmic behavior profoundly alter this scenario: the super-Ohmic regime exhibits no phase transition, whereas the sub-Ohmic regime displays a continuous second-order transition, consistent with the universality classes of the spin-boson model.
Within the Ohmic regime, real-frequency linear-response calculations reveal that the phase particle does not undergo the commonly assumed diffusive-to-localized crossover. Instead, finite resistance progressively suppresses the singular zero-frequency response, producing a {\it ballistic-to-localized} change in the dynamics. At finite frequencies, coupling to the environment generates a long-lived excitation in the charge response, which evolves into a resonance as the subgap and shunt resistances are reduced.
\end{abstract}
\maketitle

Environmental coupling is typically viewed as a source of decoherence and dissipation that suppresses quantum features \cite{Weiss2021, nitzan_chemical, Leggett_spinboson, may_molecular_systems, nitzan_chemical, nielsen_qc_qinfo, Alipour_metrology, gardiner_zoller_QuantumNoise, Lidar}. However, in many situations, memory effects play a crucial role, enabling the persistence and even the enhancement of quantum coherence, resonances, and entanglement~\cite{Breuer_RevModPhys_nonmarkovian, Rivas_nonmarkvovian, Maniscalco_nonmarkovian}. Beyond simple relaxation toward equilibrium, environmental modes may compete with coherent tunneling processes and even generate quantum phase transitions (QPTs). 
Prototypical examples are the spin‑boson model~\cite{Leggett_spinboson, Weiss2021} and the quantum Rabi model~\cite{Rabi_original, JCM_rabi}, which have attracted sustained interest due to their conceptual simplicity, experimental accessibility, and broad relevance—particularly in quantum optics~\cite{Rabi_optics1, Rabi_optics2, Rabi_optics3}.

In these paradigmatic two‑level settings, the existence of dissipation‑driven QPTs has been extensively elucidated~\cite{Giulio_spinboson_PRB, Giulio_manyspin_PRB, Grazia_NatComm, Giulio_PRL}. By contrast, the emergence of a QPT in a shunted Josephson junction—a system effectively equivalent to a massive particle moving in a periodic potential and coupled to a bath of harmonic oscillators (representing the shunt resistance)—has remained the subject of long‐standing debate~\cite{Murani_absence,  Comment_on_absence, Reply_to_comment, Altimiras:2023trq, Capone_QBM, Masuki_Absence, Masuki_reply, Comment_absence_Sepulcre}. 

In pioneering papers on this subject, the so-called dissipative phase transition has been predicted~\cite{Schmid, Bulgadaev}. At weak dissipation, macroscopic quantum tunneling delocalizes the phase (the coordinate of the massive particle) across different minima of the Josephson potential. At strong dissipation, tunneling is suppressed, and the phase particle becomes localized, leading to the restoration of a coherent Josephson response within the conventional interpretation. The transition is expected to affect the mobility of the phase particle evolving from a finite constant to zero at QPT ~\cite{Schmid, Bulgadaev}. Perturbative analyses in the presence of a small current bias suggested that, in the weak-dissipation regime, the low-frequency response is dominated by the shunt channel, making the junction an ideal insulator~\cite{Likh1, Likh2, Likh3}. Different experimental attempts~\cite{Exp_1, Exp_2, Exp_3} have been made to observe the theoretically predicted dissipation-driven phase transition in a small Josephson junction, but the interpretation of these results is still debated. Recently, even the absence of this QPT in the predicted parameter regime has been reported~\cite{Murani_absence, Masuki_Absence, Masuki_reply, Altimiras:2023trq}. The main objection stems from the observation that the Cooper‑pair box—corresponding to the limit of {\it infinite} shunt resistance—can sustain ac supercurrents, whereas in the standard dissipative scenario it would be placed deep in the insulating phase of the predicted QPT. In the new proposed scheme, QPT does not take place and the Josephson junction is predicted to be superconducting, within the linear‑response regime, irrespective of the coupling strength with the environment.

The central issue is not merely the existence of the equilibrium Schmid transition but how it emerges from a quasiparticle-based description and how it manifests itself in the real-frequency linear response. In particular, the regular DC mobility must be distinguished from possible singular zero-frequency contributions, since, in general, the imaginary-axis limit does not necessarily reproduce the physical real-frequency response.

In this Letter, we start from a quasiparticle-based effective action for the Josephson phase. We show that quantum phase fluctuations combined with Ohmic or sub-Ohmic environmental modes induce a QPT associated with localization of the phase particle. We then analyze the real-frequency linear response and demonstrate that this equilibrium transition has a direct dynamical signature: the singular ballistic contribution to the phase-particle mobility disappears at the transition, while the regular DC mobility remains zero. Our results connect the equilibrium localization transition and its transport response.

{\it The model.}
 The Josephson effect occurs at a weak electrical contact between two superconductors. In pioneering works~\cite{Josephson1, Josephson2, Feynman3}, this system was modeled as a nonlinear inductor, treating the superconducting phase difference $\varphi$ as a classical variable. Subsequent models incorporated the effects of phase fluctuations and quasiparticle (QP) tunneling. 
 A microscopic model incorporating both effects~\cite{PhysRevLett_Schon, PhysRevB_Schon} yields an effective imaginary-time action for the phase difference $\varphi(\tau)$:
\begin{equation}\begin{split}\label{eqn:action_QP}
&\mathcal{S}_{QP}[\varphi(\tau)] = \frac{\hbar^2}{4E_C}\int_0^{\beta \hbar} d \tau \biggl(\frac{d\varphi}{d\tau}\biggr)^2 -E_J\int_0^{\beta \hbar} d \tau \cos\varphi(\tau)  \\
&+ \int_0^{\beta \hbar} d \tau  \int_0^{\beta \hbar} d\tau' \, K(\tau-\tau')\sin^2\biggl( \frac{\varphi(\tau)-\varphi(\tau')}{4}\biggr).
\end{split}
\end{equation}
Here, $E_C$ is the charging energy of the junction capacitance, and $E_J = \hbar I_c/2e$ is the Josephson coupling energy, $I_c$ being the critical current. The memory kernel $K(\tau)$, which encodes the effects of QP tunneling, is defined as
\begin{equation}\label{eqn:def_kernel}
    K(\tau) = \frac{2\hbar}{\pi^2}\int_0^{+\infty} d\omega \, J(\omega) D_\omega (\tau),
\end{equation}
where $J(\omega)$ is the spectral function that captures the dissipation physics, and
\begin{equation}\label{eqn:def_bos}
D_\omega(\tau) = \frac{\cosh\bigl[\omega(\beta\hbar/2 -|\tau|)\bigr]}{\sinh(\beta\hbar\omega/2)}
\end{equation}
is the free bosonic propagator. In this Letter, we consider a phenomenological spectral function
\begin{equation}\label{eqn:def_spectral}
J(\omega) = \alpha \omega^s \omega_D^{1-s} \Theta(\omega_D - \omega),
\end{equation}
which describes Ohmic ($s=1$), sub-Ohmic ($s<1$), or super-Ohmic ($s>1$) dissipation regimes. The exponent $s$ governs the long-time asymptotic decay of the kernel. Indeed, from Eqs.~\eqref{eqn:def_kernel} and~\eqref{eqn:def_spectral}, $K(\tau) \sim \tau^{-1-s}$ for $1/\omega_D^{} \ll \tau \ll \beta\hbar/2$.

For Ohmic dissipation, $J(\omega)$ depends linearly on the frequency up to the cutoff point $\omega_D$, and the dimensionless dissipation strength $\alpha = R_q / R_{QP}$ represents the ratio of quantum resistance $R_q = h/4e^2$ to effective resistance $R_{QP}$. This quantity describes the effective parallel combination of shunt and subgap resistances \cite{PhysRevLett_Schon, PhysRevB_Schon,Guinea_Schon_jap}.
We emphasize that the trigonometric dependence on the phase difference in Eq.~\eqref{eqn:action_QP} reflects the inherently discrete nature of single-electron tunneling across the insulating barrier. Expanding the sine term to the lowest non-trivial order formally recovers the phenomenological Caldeira-Leggett (CL) model for an Ohmic shunt, corresponding to a continuous charge flow. In the following, we will focus on the equilibrium properties of the system when either Ohmic or non-Ohmic dissipation occurs.

The action in Eq.~\eqref{eqn:action_QP} can be derived from the Hamiltonian of a Josephson junction, characterized by its capacitance $C = 2e^2/E_C$ and the phase-related flux $\phi = \frac{\phi_0}{2\pi}\varphi$ ($\phi_0=h/2e$ being the flux quantum), coupled to two independent, infinite sets of LC circuits, i.e., bosonic baths. The junction interacts with these two environmental baths via the operators $\cos\big(\pi\phi/\phi_0\big)$ and $\sin\big(\pi\phi/\phi_0\big)$, respectively, with coupling constants completely described by the spectral function $J(\omega)$. For the explicit Hamiltonian, see the Supplemental Material~\cite{SuppMat}. The cutoff frequency $\omega_D$ is associated with the highest energy scale in the problem, related to the Fermi energy of the two superconductors. Throughout this work, we adopt $E_C$ as the reference energy scale and fix $\hbar\omega_D = 50E_C$ and $E_J/E_C=0.5$.

{\it Thermodynamics and equilibrium properties.}
Here, we investigate the equilibrium effects induced by quantum phase fluctuations and resistances using a Worldline Monte Carlo (WLMC) approach. Note that the dissipative term in Eq.~\eqref{eqn:action_QP} mediates a long-range interaction with periodicity $4\pi$. Indeed, since the kernel multiplies
$\sin^2\biggl(\frac{\varphi(\tau)-\varphi(\tau')}{4}\biggr)$, two portions of a path differing by $4\pi$ give no dissipative cost, while portions differing by $2\pi$ maximize it. At zero temperature and sufficiently large coupling $\alpha$, we therefore expect the phase particle to undergo a QPT, localizing in either the even or odd minima of the Josephson potential, provided that $K(\tau)$ decays sufficiently slowly.

\begin{figure}[]
\centering
\includegraphics[width=\columnwidth]{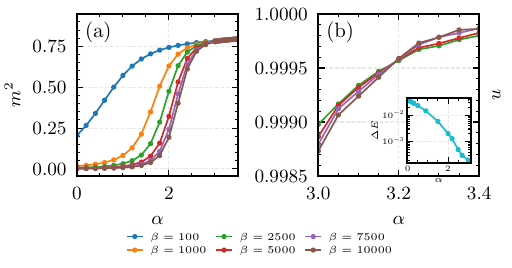}
\caption{Order parameter $m^2$ as a function of $\alpha$ for different values of $\beta$ (in units of $1/E_C$), in the Ohmic case ($s=1$). The order parameter becomes non-zero as the coupling $\alpha$ is increased (a). Binder cumulant $u$ for the same values of $\beta$, showing a scale-invariant crossing at $\alpha_c \approx 3.2$ (b). Inset: Bandwidth $\Delta E$ as a function of $\alpha$, obtained via DMRG.}
\label{fig:QPT}
\end{figure}

To identify the critical coupling, we introduce the order parameter
\begin{equation}\label{eqn:order_parameter_qpt}
m^2 = \frac{1}{\beta\hbar}\int_0^{\beta\hbar} d\tau \,\bigl\langle \cos\bigl(\varphi(\tau)/2\bigr)\cos\bigl(\varphi(0)/2\bigr)\bigr \rangle.    
\end{equation}
This quantity vanishes in the symmetric phase and becomes finite once symmetry is broken, where the system selects either the even or odd minima of the Josephson potential. We also define the Binder cumulant $u = \frac{3}{2} - \langle m^4\rangle /( 2\langle m^2\rangle^2)$. As illustrated in Fig.~\ref{fig:QPT}, the QPT is unambiguously signaled by both the onset of a finite $m^2$ and a clear crossing point in $u$, which yields a critical coupling $\alpha_c \approx 3.2$. Above this threshold, the phase particle remains localized in even or odd minima. The effective bandwidth for $\alpha \ge \alpha_c$ has long been expected to be extremely small \cite{korshunov}. This behavior is clearly confirmed by our density‑matrix renormalization group analysis (DMRG) \cite{SuppMat}, as illustrated in the inset of Fig.~\ref{fig:QPT}b. Conversely, because the bandwidth collapses, finite-temperature effects become indistinguishable from full localization in one of the minima.

The next step is to clarify the nature of the QPT. Crucially, a straightforward expansion of the term $\sin^2\biggl( \frac{\varphi(\tau)-\varphi(\tau')}{4}\biggr)$ in Eq.~\eqref{eqn:action_QP} reveals that the action can be exactly decomposed into the unshunted junction contribution and two independent non-local terms--mediated by the memory kernel $K(\tau-\tau')$--acting, respectively, on $\cos(\varphi/2)$ and $\sin(\varphi/2)$. When the phase particle is located near the minima of the Josephson potential, $\sin(\varphi/2)$ vanishes, leaving the QPT to be entirely governed by the cosine interaction. In this regime, $\cos(\varphi/2) \approx \pm 1$ acts as a discrete pseudospin variable. This establishes a natural low-energy mapping to a spin-boson model~\cite{Leggett_spinboson}, where the effective two-level system distinguishes the two $4\pi$-periodic families of Josephson minima.

As a consequence, we expect that the two models share the same universality class. Then, we analyze the Ohmic ($s=1$) and sub-Ohmic ($0<s<1$) regimes using a Berezinskii-Kosterlitz-Thouless (BKT)~\cite{Giulio_PRL, Giulio_manyspin_PRB, Giulio_spinboson_PRB} and a second-order \cite{Luijten_crit, Bulla_spinboson} scaling framework, respectively. 
\begin{figure}[]
\centering
\includegraphics[width=\columnwidth]{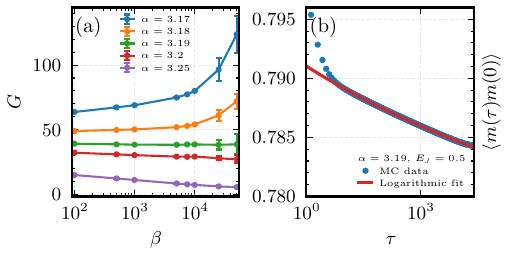}
\caption{Scaling function $G$ as a function of $\beta$ for different values of $\alpha$ ($s=1$). $G$ turns out to be independent of $\beta$ at $\alpha \approx 3.19$ (a). Correlation function $\langle m(\tau)m(0)\rangle$ (blue dots) and logarithmic fit $0.7609(5) + 1.03(4)/(\ln \tau + 34.2(8))$ (red line) as functions of $\tau$ (in units of $\hbar/E_C$) at $\alpha = 3.19$ (b).}
\label{fig:BKT}
\end{figure}
For the Ohmic case, the long-time expansion of the transition-driving term yields $\mathcal{S}_{\mathrm{crit}}[\varphi(\tau)] \simeq -\frac{1}{2}\int d\tau d\tau' \cos\bigl[\varphi(\tau)/2\bigr] \frac{\alpha_{\mathrm{eff}}}{2\tau^2} \cos\bigl[\varphi(\tau')/2\bigr]$, with $\alpha_{\mathrm{eff}} = 4\hbar \alpha/\pi^2$. Defining the scaling function $\Psi(\alpha,\beta ) = \alpha_{\text{eff}}m^2$, the BKT theory dictates $\Psi(\alpha_c, \beta)/\Psi_{c} = 1+ [2(\ln\beta-\ln \beta_0)]^{-1}$, with $\Psi_c = 1$ \cite{Minn_PRB1, Minn_PRL, Minn_PRB2}. Consequently, the quantity
\begin{equation}\label{eqn:scaling_G}
G(\alpha, \beta) \equiv \frac{1}{\Psi(\alpha, \beta)-1}-2\ln\beta = \frac{1}{4\hbar \alpha m ^2/\pi^2 -1} - 2\ln \beta    
\end{equation}
must become temperature-independent at criticality \cite{Grazia_NatComm, Giulio_PRL, Giulio_manyspin_PRB, Giulio_spinboson_PRB, Capone_QBM}. As shown in Fig.~\ref{fig:BKT}(a), this criterion yields $\alpha_c \approx 3.19$, which is highly consistent with the critical point extracted from the Binder cumulant crossing. At this critical point, we also observe the expected logarithmic decay $A + B/(\ln\tau + C)$ of the correlation function $\langle m(\tau)m(0)\rangle$\cite{Bhattacharjee_log_corr, Luijten_log_corr, Capone_QBM} (Fig.~\ref{fig:BKT}(b)), where $m(\tau) \equiv \cos(\varphi(\tau)/2)$.

Extending our numerical analysis to the non-Ohmic regime ($s \neq 1$), we probe the expected second-order QPT for sub-Ohmic dissipation ($s<1$) via the Binder cumulant $u$ (see Appendix A). On the other hand, in the super-Ohmic regime ($s>1$), the crossings of the Binder parameter $u$ at different temperatures do not define a stable fixed point, revealing an absence of critical behavior at zero temperature (see Appendix A). All these findings confirm that the equilibrium properties of the system fall entirely within the universality class of the spin-boson model. We emphasize the key differences with respect to the CL model: (i) In the Ohmic regime, the critical coupling $\alpha_c$ depends strongly on the ratio $E_J/E_C$ (e.g., $\alpha_c \approx 3.3$ and $3.5$ for $E_J/E_C = 0.4$ and $0.3$, respectively), while in the CL model, $\alpha_c = 1$ regardless of $E_J/E_C$. (ii) Unlike the CL model, where only $s=1$ induces a QPT, the criticality in the QP model is not fragile and persists in the sub-Ohmic regime ($s<1$).  

{\it Linear Response Theory.}
Linear response theory provides the standard criterion for discriminating between insulators, metals, and superconductors. To characterize the transport properties of the junction, we define the response function $\Psi_A(z)$ for a generic observable $A$ as~\cite{Scalapino, Evertz_PRB, Shastry_PRB}
\begin{equation}
\Psi_A(z) = \frac{i}{z}\biggl[\Pi_A(z)+\langle k\rangle\biggr],   
\end{equation}
where $\Pi_A(z)$ is the Fourier transform of the correlation function associated with $A$, and $k = \frac{i}{\hbar}[A,B]$, with $B$ being the operator such that $dB/dt = A(t)$. Here, $z = \omega + i\epsilon$ (with $\epsilon>0$) is a complex variable in the upper half-plane.

On the real axis ($z \to \omega + i 0^+$), the real part of the response yields:
\begin{equation}
\mathrm{Re} \,\Psi_A(\omega) = \pi D_A\delta(\omega) + \Psi_{A,\text{reg}}(\omega),
\end{equation}
where the even function $\Psi_{A,\text{reg}}(\omega)$ is the regular part of $\Psi$, and $D_A$ is the Drude weight. 

Introducing the Matsubara Green function $\Pi_A(\tau) = -\frac{1}{\hbar} \bigl\langle T_{\tau} A(\tau)A(0)\bigr\rangle$ and its Fourier coefficients $\Pi_A(i\omega_n)$ with $\omega_n = 2\pi n/\beta \hbar$, the Drude weight can be rigorously evaluated as:
\begin{equation}\label{eqn:Drudea}
    D_A = \langle k \rangle + \Pi_{A}(i\omega_n \to 0).
\end{equation}
A property closely related to the Drude weight is the Meissner stiffness, defined by: 
\begin{equation}\label{eqn:Drudea1}
    D_{A,M} = \langle k \rangle + \Pi_{A}(i\omega_n =0).
\end{equation}
Their difference comes from transitions within degenerate manifolds \cite{Evertz_PRB, Shastry_PRB}: $D_{A,M}-D_A = -\beta \sum_{E_n=E_m} p_n|\bigl \langle n| A | m \bigr \rangle|^2$, where $p_n$ is the Boltzmann weight of the eigenstate $|n\rangle$. The full derivation of these relations, as well as their connection to the Mori relaxation function, are detailed in Appendix~B. 

Under a current bias $I$ (see Appendix~B), the dimensionless mobility $\mu(z)$ of the phase particle acts as an effective impedance of the circuit. It is related to the response function $\Pi_Q$ by:
\begin{equation}\label{eqn:mob1}
    \mu(z)=\frac{1}{R_{QP} C^2}\Psi_Q(z),
\end{equation}
i.e., in this bias scheme, the operator $A$ represents the charge $Q$ and $\langle k \rangle$ turns out to be a constant, $C$. The physical response of the system is therefore fully encoded in the real-frequency function, $\mathrm{Re}\, \mu(\omega) \equiv \lim_{\epsilon \to 0^+} \mathrm{Re}\, \mu(\omega + i\epsilon)$.

\begin{figure}[]
\centering
\includegraphics[width=\columnwidth]{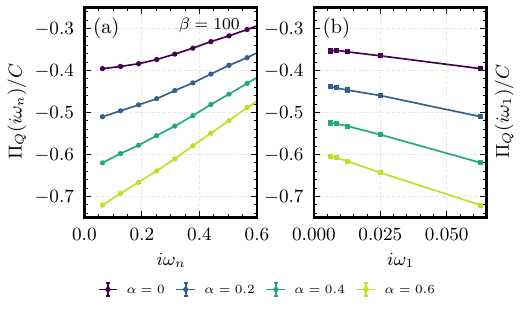}
\caption{(a) Normalized charge-charge correlation function $\Pi_Q(i\omega_n)/C$ (with $C = \langle k\rangle$) versus Matsubara frequency $i\omega_n$ at inverse temperature $\beta E_C= 100$ for various $\alpha$. (b) The same normalized function evaluated at the first Matsubara frequency, $i\omega_1 = 2\pi/\beta$ (in units of $E_C/\hbar$), versus $i\omega_1$ for different $\alpha$. For $\alpha < \alpha_c$, the ratio does not converge to $-1$, implying a non-vanishing $D_Q$ also for $\alpha \ne 0$.}
\label{fig:pi_q}
\end{figure}

First, we emphasize the singular behavior of $\Psi_Q(z)$ when $\alpha=0$. In the Cooper pair box, the exact diagonalization shows that $\mathrm{Re} \,\Psi_Q(\omega)$ exhibits a singularity at $\omega = 0$, with $D_Q$ remaining finite. In particular, $\Pi_{Q}(i\omega_n =0)=-\langle k \rangle$, while $\Pi_{Q}(i\omega_n \to 0) \ne -\langle k \rangle $ (see also Fig.~\ref{fig:pi_q}), implying $D_{Q,M}=0$ but $D_Q \ne 0$. This inequality demonstrates that, within linear response, the Cooper-pair box behaves as an insulator, characterized by a dissipationless dynamics of the phase particle. One is therefore forced to go beyond linear response, where Bloch oscillations—allowing Cooper pairs to tunnel through the junction—are recovered. 

The key question is how this scenario is modified once QP resistance is included. To this end, we recall that in the mechanical analog, $I$, $V$, and $C$ play the roles of the electric field, velocity, and mass, respectively. Here, one expects that at any non vanishing temperature, the delta function, present at $\alpha=0$ (free electrons), becomes a Lorentzian, i.e. a Drude peak appears in the spectrum, due to the scattering with bosons, so that the behavior of the fluctuations switches from ballistic to diffusive, by turning on the electron-phonon coupling. This is the typical scenario encountered in polaron physics \cite{Mahan,SuppMat}, and it is the underlying principle of the theory proposed by Schmidt in the weak-coupling regime. Since $D_Q$ suddenly vanishes by switching on $\alpha$, within this scenario $\Pi_{Q}(i \omega_n \to 0)$ turns out to be a non analytic function of $\alpha$ \cite{SuppMat}: $\Pi_{Q}(i \omega_n \to 0)=-\langle k \rangle$ as soon as $\alpha \ne 0$. This non-analyticity leads to the well‐known challenges associated with the calculation of transport properties in metals.

A natural question is whether the proposed scenario is consistent with the WLMC data. Fig.~\ref{fig:pi_q} shows $\Pi_{Q}(i \omega_n) / \langle k \rangle$ as a function of $\alpha$ and $\beta$ in the QP model. The plots clearly prove that $\Pi_{Q}(i \omega_n)$ is analytic in $\alpha$: the ratio $\Pi_{Q}(i \omega_n \to 0) / \langle k \rangle$ does not approach $-1$ by switching on $\alpha$. As a consequence, the proposed scenario must be substantially revised.

\begin{figure}[]
\centering
\includegraphics[width=\columnwidth]{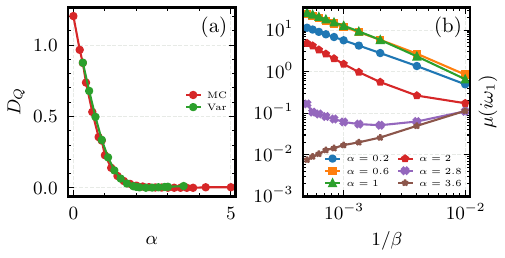}
\caption{Singular contribution $D_Q$ (in units of $e^2/E_C$) to the response function $\mu(\omega)$ as a function of $\alpha$ at $\beta = 100$ (in units of $1/E_C$), computed via Monte Carlo (red dots) and variational approach (green dots) (a). Dimensionless mobility $\mu(i\omega_1)$ at first imaginary frequency as a function of $1/\beta$ (b).}
\label{fig:dq_mob}
\end{figure}

We begin by determining $D_Q$ from the Matsubara correlation function $\Pi_Q$, following the method introduced in \cite{Evertz_PRB}. Then we compute the dynamical spectra $\Psi_{Q,\text{reg}}(\omega)$ using the Maximum Entropy method (MaxEnt) \cite{jarrell} starting from the numerically exact WLMC data on the imaginary axis (see also \cite{SuppMat}). As shown in Fig.~\ref{fig:dq_mob}(a), $D_Q$ remains finite in the delocalized phase but vanishes in the symmetry-broken phase. Note that $D_Q$ is also successfully compared with results obtained using a variational approach \textit{à la Feynman} \cite{SuppMat}. 

We now turn to another crucial point that emerges from the standard approach commonly used in the literature. Dimensionless mobility $\mu(z)=\frac{V(z)}{I(z) R_{QP}}$ ($z \to 0$) has been predicted to approach $1$ as $\alpha \to 0$ and vanish above $\alpha_c$ \cite{Likh1, Likh2, Likh3}. As a consequence, the entire bias current flows through the resistance, and the junction behaves as an ideal insulator. This prediction has been formulated by examining the behavior of $\mu(i\omega_n \to 0)$ along the imaginary axis in the CL model of the junction. Fig.~\ref{fig:dq_mob}(b) shows the mobility at the first imaginary frequency in the QP model. The plots clearly show that it does not approach 1 at small values of $\alpha$. 

In any case, the main criticism of this theory stems from the observation that the DC mobility, $\mu(\omega \to 0)$, has been evaluated by performing the limit along the imaginary axis. This procedure is well defined only when there are no singularities at $z = 0$, the origin of the complex plane. Since $D_Q$ remains finite in the delocalized phase (Fig.~\ref{fig:dq_mob}(a)), $\mu(z)$ has a zero-frequency singularity and the limits $i\omega_n\to 0$ and $\omega\to 0^+$ provide two different results. Therefore, an imaginary-axis extrapolation incorrectly attributes the singular Drude-like contribution to a finite DC mobility. 

\begin{figure}[]
\centering
\includegraphics[width=\columnwidth]{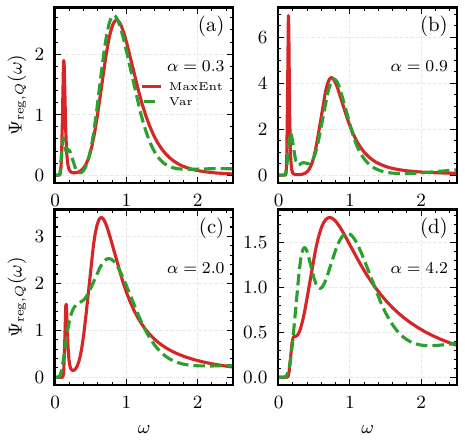}
\caption{Analytic reconstruction of the regular part of the response function $\Psi_{\mathrm{reg},Q}$ (in units of $e^2 \hbar/E_C^2$) at $\beta = 100$ (in units of $1/E_C$) as a function of the frequency $\omega$ (in units of $E_C/\hbar$). Red (green) curves show the results obtained via MaxEnt algorithm (variational approach \cite{SuppMat}).}
\label{fig:an_ric}
\end{figure}

In Fig.~\ref{fig:an_ric}, we explicitly plot the analytically continued regular part of the charge spectral function in the Ohmic regime. Mobility $\mu_{\mathrm{DC}}$ vanishes for all  couplings $\alpha$, whereas, below the critical point ($\alpha<\alpha_c$), a zero-frequency delta function with weight $D_Q \neq 0$ persists, indicating that the whole system behaves as a strict insulator within the linear response regime. Contrary to common expectations, the system does not conduct DC current resistively (it would imply finite mobility and the presence of a Drude-like term!). As a consequence, any applied DC current bias is entirely converted into an AC signal, pointing to nonlinear transport effects fundamentally akin to those of an isolated Cooper-pair box. In contrast, above the transition ($\alpha>\alpha_c$), $\mu_{\mathrm{DC}}$ vanish and $D_Q$ is extremely small, being rigorously zero at $T=0$ \cite{SuppMat}. It signals the disappearance of the ballistic channel and the onset of a state capable of sustaining dissipationless DC transport.  
This intriguing behavior of the DC current in the linear‑response regime leads to a ballistic‑to‑localized transition in the phase‑particle dynamics \cite{SuppMat}. Finally, at finite frequency, in the weak-coupling regime, a well-defined excitation appears, at energies around the bare bandwidth. It broadens into a resonance as the bath coupling increases and is ultimately suppressed near $\alpha_c$ due to enhanced scattering with bosonic excitations. Note that, at $\alpha = 0$, only the highest energy peak, located at the plasma frequency $\hbar\omega_p \approx \sqrt{2E_J E_C}$, is observed.

{\it Conclusions.}
We have established that the equilibrium thermodynamics of the shunted Josephson junction falls entirely within the spin-boson universality class. Furthermore, our real-frequency linear-response analysis demonstrates that extracting DC transport properties from the imaginary axis is fundamentally flawed. Rather than exhibiting a resistive regime, exact mobility reveals that the system undergoes a QPT from a strict insulator to an ideal superconductor. Interestingly, below the critical coupling, no DC current flows resistively through the shunt; instead, any applied DC bias is completely converted into an AC signal. Our results demonstrate that dissipation-induced QPTs in Josephson circuits must be identified through their real-frequency dynamical signatures, rather than through imaginary-axis extrapolations. This insight reshapes the theoretical foundations of dissipative superconducting devices and highlights the essential role of nonlinear effects.

{\it Acknowledgments.}
G.D.F., R.F., C.A.P, and A.d.C. acknowledge financial support from PNRR MUR Project No. PE0000023-NQSTI. C.A.P. and G.D.B acknowledges funding from IQARO (Spin-orbitronic Quantum Bits in Reconfigurable 2DOxides) project of the European Union’s Horizon Europe research and innovation programme under grant agreement n. 101115190. R.F. acknowledges funding from the European Research Council (ERC), Grant agreement No. 101053159 – RAVE. N.N. was supported by JSPS KAKENHI Grant Numbers 24H00197, 24H02231 and 24K00583. N.N. was supported by the RIKEN TRIP initiative.

\bibliography{main_text}

\clearpage

{\Large \bf End Matter}
\appendix
\section{Appendix A: Presence (absence) of QPT in the sub-(super-)Ohmic regime}

\begin{figure}[]
\centering
\includegraphics[width=\columnwidth]{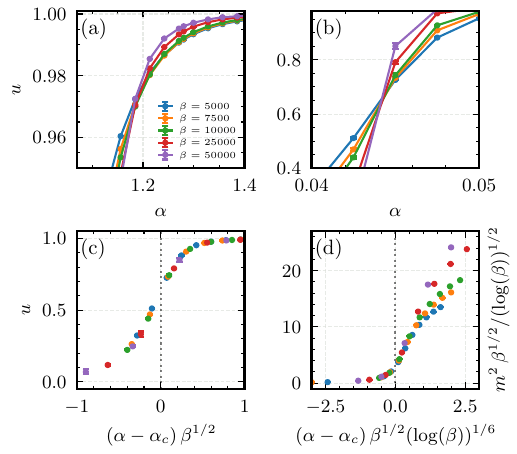}
\caption{Binder parameter $u$, for different values of $\beta$, in the sub-Ohmic regime $s = 0.9$ (a) and $s = 0.5$ (b), whose crossing occurs at $\alpha_c \approx 1.175$ and $\alpha_c \approx 0.44$ respectively. Data collapse at $s = 0.5$ is consistent with the mean field critical exponents $\nu = 2$ (c) and $\gamma = 1$ (d). The collapse improves with
increasing $\beta$ on the localized side $\alpha>\alpha_c$. A slow convergence is expected at the upper critical dimension $s=0.5$, where logarithmic corrections are relevant \cite{Bulla_spinboson}.}.
\label{fig:subohmic}
\end{figure}

Here we extend our numerical analysis to the non-Ohmic regime ($s \neq 1$), where we probe the expected second-order QPT for sub-Ohmic dissipation ($s<1$) via the Binder cumulant $u = \frac{3}{2} - \langle m^4\rangle /( 2\langle m^2\rangle^2)$. The robustness of the QPT is certified by a clear fixed point in $u$ at finite $\alpha_c$ for both $s = 0.9$ and $s = 0.5$ (Figs.~\ref{fig:subohmic}(a,b)). At the upper critical dimension $s = 0.5$, the mean-field exponents $\nu = 2$ and $\gamma = 1$ are anticipated, and the observables should satisfy the scaling relations~\cite{Luijten_crit, Bulla_spinboson}:
\begin{subequations}
\begin{align}
    u &= f\bigl((\alpha-\alpha_c) \beta^{1/\nu} \bigr),\\
    \beta^{1-\gamma/\nu}m^2 (\ln \beta)^{-\mu} &= g\bigl((\alpha -\alpha_c) \beta^{1/\nu} (\ln \beta)^{\lambda} \bigr),
\end{align}
\end{subequations}
where $f(x)$ and $g(x)$ are universal functions. The parameters $\mu = 1/2$ and $\lambda = 1/6$ account for logarithmic corrections that cancel residual finite-size dependencies at finite $\beta$. Figs.~\ref{fig:subohmic}(c) and~\ref{fig:subohmic}(d) validate these relations through a data collapse analysis.
\begin{figure}[]
\centering
\includegraphics[width=\columnwidth]{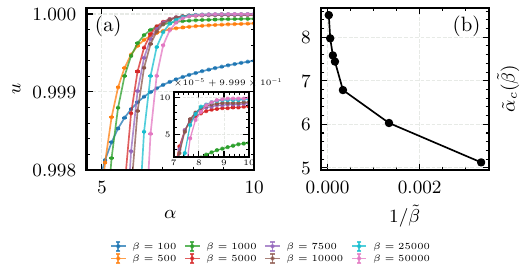}
\caption{Binder parameter $u$ for different values of $\beta$, in the super-Ohmic regime ($s = 1.1$) in the range of $\alpha$ where the curves intersect each other (a). Critical coupling ${\tilde \alpha_c}$ as a function of $1/{\tilde \beta}$ in the super-Ohmic regime ($s = 1.1$), which evidences the tendency to diverge of ${\tilde \alpha_c}$ as the temperature is lowered (b).}
\label{fig:s1.1}
\end{figure}

 In the super-Ohmic regime ($s>1$), a simple argument precludes a finite-$\alpha$ QPT. If the particle were to localize, the free energy cost $F_{2\pi}$ associated with a phase slip between adjacent minima would need to be positive. However, the action cost for a $2\pi$ phase slip in the interval $\tau_1<\tau<\tau_2$ is bounded by ${\cal S}_{2\pi} = \int d\tau \,\tau K(\tau) = \alpha C$ (with $C$ constant for $s>1$ as $\beta \to \infty$), while the gain of entropy is $S_{\mathrm{ent}} \propto  2\ln(\beta \hbar)$. Consequently, $F_{2\pi} = {\cal S}_{2\pi} - S_{\mathrm{ent}} \to -\infty$ for $\beta \to \infty$, demonstrating that such phase slips are not penalized and the system is always fully delocalized. We confirmed this numerically for $s = 1.1$ (just above the Ohmic threshold). As shown in the following, the crossings of the Binder parameter $u$ at different temperatures do not define a stable fixed point, revealing the absence of critical behavior at zero temperature. Indeed, in the presence of a critical point, the curves of $u(\alpha, \beta)$ for different $\beta$ are expected to intersect at a stable fixed point $\alpha_c$, with the intersection becoming progressively sharper as $\beta$ increases. We define effective critical coupling $\tilde{\alpha}_c(\beta_1, \beta_2)$ as the intersection point of two curves with average inverse temperature $\tilde{\beta} = (\beta_1 + \beta_2)/2$. In a critical system, $\tilde{\alpha}_c(\tilde{\beta})$ should converge to $\alpha_c$ as $\tilde{\beta} \to \infty$. In contrast, a divergence of $\tilde{\alpha}_c$ as $\tilde{\beta} \to \infty$ indicates the absence of a transition. As shown in Fig.~\ref{fig:s1.1} , for $s = 1.1$, i.e., slightly above the Ohmic regime, $\tilde{\alpha}_c$ diverges as the temperature is reduced, confirming the absence of critical behavior.

\section{Appendix B: Linear transport, voltage and current bias}{}\label{sec:app_transport}
Given an observable $A$ and an operator $B$ such that $\frac{dB}{dt} =\frac{1}{i\hbar} [B,H] = A(t)$, the response function $\Psi_A(z)$ can be expressed via the Mori relaxation function framework. The response function is given by:
\begin{equation}
\Psi_A(z) = \frac{i}{z}\biggl[\Pi_A(z)+\langle k\rangle\biggr],   
\end{equation}
where $\Pi_A(z)$ is the Fourier transform of the correlation function $\Pi_A(t-t')=-\frac{i}{\hbar}\theta(t-t')\bigl\langle[A(t),A(t')] \bigr\rangle$ associated with $A$, and $k = \frac{i}{\hbar}[A,B]$. It is well known that $\Pi_A(z)$ is analytic in the upper complex half-plane. By performing the spectral decomposition in the exact eigenstate basis, the regular part $\Psi_{A,\text{reg}}(\omega)$ is written as:
\begin{equation}\label{eq_app:decomposition_regular}
    \begin{split}
        &\Psi_{A,\text{reg}}(\omega)=\\  &\pi\sum_{\substack{n,m \\ E_n\neq E_m}}\frac{\left|A_{nm}\right|^2}{E_m-E_n}\frac{e^{-\beta E_n} - e^{-\beta E_m}}{Z_p}\delta\biggl(\omega - \frac{E_m-E_n}{\hbar}\biggr).
    \end{split}
\end{equation}
Analogously, analyzing the limit $z \to 0$ in the upper half-plane ($\epsilon \to 0^+$) of $\Psi_A(z)$ \cite{Evertz_PRB, Shastry_PRB}, one recovers the expression for the Drude weight $D_A = \langle k \rangle + \Pi_{A}(i\omega_n \to 0)$ derived in Eq.~\eqref{eqn:Drudea} of the main text. Again, evaluating the static Matsubara response $\Pi_{A}(i\omega_n =0)$ directly yields the Meissner stiffness $D_{A,M}$ (Eq.~\eqref{eqn:Drudea1}).
The difference $D_{A,M} - D_A$ captures the exact zero-frequency limits arising from degenerate states $E_n = E_m$, which leads directly to:
\begin{equation}
    D_{A,M}-D_A = -\beta \sum_{E_n=E_m} p_n|\bigl \langle n| A | m \bigr \rangle|^2.
\end{equation}

In what follows, we detail both the voltage and current bias schemes.

Let us first introduce the Heisenberg representation for the current $I(t) = -\dot{Q}$ and the voltage $V(t) = \dot{\phi} = Q/C$. A voltage bias $\delta V(t)$ applied across the junction for $t>0$ introduces a small perturbation $I \delta\phi(t)$ into the Hamiltonian, with $\delta V(t) = \frac{d}{dt} \delta \phi(t)$ driven by a classical time-dependent field $\delta \phi(t)$. As customary, the system is assumed to reside in thermodynamic equilibrium for $t<0$. At the first order in this perturbation, the total current across the junction reads $I_{\text{tot}}(t) = I(t) + \frac{\partial I}{\partial \phi}\delta \phi(t)$. Linear response theory \cite{Mahan} then dictates that the expectation value of the total current variation is
\begin{equation}\label{eqn:linear_current_tot}
\delta \langle I_{\text{tot}}\rangle (t)  = \int_{-\infty}^{+\infty}dt' \,\Pi_I(t-t')\delta\phi(t') + \biggl\langle\frac{\partial I}{\partial \phi}\biggr\rangle\delta\phi(t).
\end{equation}
By introducing the Fourier transform of the linear admittance, $Y(z) = \int_0^{+\infty}dt \,Y(t)e^{izt} = \delta\langle I_{\text{tot}}\rangle(z)/\delta V(z)$, one can readily show from Eq.~\eqref{eqn:linear_current_tot} that
\begin{equation}\label{eqn:kubo_current}
Y(z) = \frac{i}{z}\biggl[\Pi_I(z)+\biggl\langle\frac{\partial I}{\partial\phi}\biggr\rangle\biggr],   
\end{equation}
where the complex frequency $z = \omega +i\epsilon$ (with $\epsilon >0$) resides in the upper half-plane. 

Equation~\eqref{eqn:kubo_current} directly links the admittance to the Kubo-Mori relaxation function evaluated for the current operator $A=I$, such that $\Psi_{I}(z)=Y(z)$. The limit $\epsilon \to 0^+$ yields the real part of the response function \cite{Shastry_PRB, Evertz_PRB}:
\begin{equation}\label{eqn:kubo_tot}
\mathrm{Re} \bigl[\Psi_I(\omega)\bigr] = \pi D_I\delta(\omega) + \Psi_{I,\text{reg}}(\omega).
\end{equation}

Next, we consider the current-bias scheme, which naturally involves the charge operator $A=Q$ and its corresponding response function $\Psi_Q(z)$. We begin by incorporating a small current bias $-I(t)\phi$ into the Hamiltonian, resulting in $H_{\text{tot}}=H-I(t)\phi$. Here, $I(t)$ is a prescribed time-dependent profile for $t>0$ (with equilibrium assumed for $t<0$). Applying a gauge transformation $\chi(t)=e^{-S}\chi^{\prime}(t)$ with $S=-i \frac{\phi}{\hbar}f(t)$ and $f(t)=\int I(t) dt$, the original Schr\"odinger equation $i\hbar\frac{\partial\chi}{\partial t}=H_{\text{tot}}\chi(t)$ is mapped onto $i\hbar\frac{\partial\chi^{\prime}}{\partial t}=H^{\prime}_{\text{tot}}\chi^{\prime}(t)$. The transformed Hamiltonian $H^{\prime}_{\text{tot}}$ corresponds to $H$ with the charging energy $\frac{Q^2}{2C}$ substituted by $\frac{(Q+f(t))^2}{2C}$. 

Effectively, this gauge shift eliminates the explicit $-I(t)\phi$ term, and the charge operator $Q$ is redefined as $\tilde{Q} = Q+f(t)$. Proceeding with standard textbook methods \cite{Mahan, Fetter}, the application of linear response theory to $\tilde{Q}$ gives the following relation in Fourier space
\begin{equation}\label{eqn:qtz}
\tilde{Q}(z)=\frac{I(z)}{C}\frac{i}{z}\bigl(C+\Pi_Q(z)\bigr).
\end{equation}
Equation~\eqref{eqn:qtz} offers an alternative perspective on the transport characteristics of the junction. By expressing the system's dimensionless impedance as $\mu(z)=\frac{1}{R}\frac{V(z)}{I(z)}$ (the impedance of the circuit is strictly related to the mobility of the phase particle \cite{Bulgadaev}), and utilizing the relation $V(z) = \frac{\tilde{Q}(z)}{C}$, we find
\begin{equation}\label{eqn:mob}
    \mu(z)=\frac{1}{R C^2}\frac{i}{z}\bigl(C+\Pi_Q(z)\bigr).
\end{equation}
This result establishes a direct equivalence between the impedance of the circuit and the charge-based Kubo-Mori function: $RC^2\mu(z) = \Psi_Q(z)$. Alternatively, dimensionless mobility can be recast via the Fourier transform of the phase-phase correlation function:
\begin{equation}\label{eqn:corrphi}
\Pi_\varphi(z)=\int_0^{\infty}dt \, e^{i z t} \Pi_\varphi(t).
\end{equation}
Double integration by parts in Eq.~\eqref{eqn:corrphi} straightforwardly demonstrates that 
\begin{equation}\label{eqn:rel}
\Pi_\varphi(z)=\biggl(\frac{2\pi}{\phi_0 z}\biggr)^2\biggl(\frac{1}{C}+\frac{1}{C^2}\Pi_Q(z)\biggr).
\end{equation}
Combining Eqs.~\eqref{eqn:mob} and \eqref{eqn:rel} then produces
\begin{equation}\label{eqn:relfin}
    \mu(z)=\frac{\alpha}{2\pi} i z \hbar \Pi_\varphi(z),
\end{equation}
thereby linking the mobility of the phase particle to $\Pi_\varphi(z)$, which can be readily evaluated within the MC approach and yields the behavior displayed in Fig.~\ref{fig:dq_mob}.

It is important to note that these exact relations simply constitute alternative representations of the phase particle's dimensionless mobility within the complex plane. However, the physical response is recovered on the real axis via $\mu(\omega)=\lim_{\epsilon \to 0^+}\mu(\omega+i\epsilon)$. Applying this limit to Eq.~\eqref{eqn:mob} leads to 
\begin{equation}\label{eqn:mobreal}
    \mu(\omega)=\frac{1}{R C^2}\bigl(\pi D_Q \delta(\omega)+\Psi_{Q,\text{reg}}(\omega)\bigr).
\end{equation}
Consequently, DC mobility, $\mu_{\mathrm{DC}}=\lim_{\omega\to 0^+}\mu(\omega)$, is dictated by the regular part of the charge response function, $\Psi_{Q,\text{reg}}(\omega)$. Notably, in the current-bias scenario, a non-zero $D_Q$ (signaled by the $\omega=0$ delta function) classifies the system as an insulator. 

Having access to $\Psi_{Q,\text{reg}}(\omega)$, $D_Q$, and $D_{Q,M}$ enables the evaluation of the real-time quantum fluctuations of the phase operator, which directly quantifies the instantaneous flux diffusivity:
\begin{equation}\label{eqn:dynamic_general1}
    \begin{split}
        &\frac{d}{dt}\biggl\langle\bigl(\phi(t)-\phi(0)\bigr)^2\biggr\rangle = \frac{2(D_Q-D_{Q,M})t}{\beta C^2} \\
        &+ \int_0^{+\infty} d\omega \, \frac{2\hbar\Psi_{Q,\text{reg}}(\omega) \sin(\omega t)}{\pi C^2\tanh(\beta\hbar\omega/2)}.
    \end{split}
\end{equation}

We emphasize that, in the case of current bias, $D_{Q,M}$ turns out to be zero, being related to the superconducting density that is vanishing in the case of a single junction (it can become different from zero for an array of Josephson junctions). It is also worth noting that Eq.~\eqref{eqn:dynamic_general1} substantiates the claim that, whenever $D_Q \ne 0$ ($D_Q=0$), the phase‑particle dynamics is ballistic (diffusive).

\clearpage

\onecolumngrid
\begin{center}
\textbf{\large Supplemental Material for\\[0.5em] ``Ballistic‑to‑Localized Dynamics as Signature of Quantum Phase Transition in Josephson Junction''}\\[1em]
\end{center}

\setcounter{equation}{0}
\renewcommand{\theequation}{S\arabic{equation}}
\setcounter{figure}{0}
\renewcommand{\thefigure}{S\arabic{figure}}
\setcounter{section}{0}
\renewcommand{\thesection}{S\Alph{section}}

\section{The Hamiltonian} The action in Eq.~(1) of the main text can be microscopically derived from the Hamiltonian:
\begin{equation}\label{eqn:hamiltonian}
\begin{split}
            &H = \frac{Q^2}{2C} - E_J\cos\biggl(2\pi\frac{\phi}{\phi_0}\biggr) \\
            &+ \sum_{j=1}^2 \sum_{i=1}^N \biggl[\frac{q^2_{i,j}}{2C_{i,j}}+\frac{\phi^2_{i,j}}{2L_{i,j}}\biggr]  \\
             &+ \sum_{i=1}^N S_{i,1}\phi_{i,1}\cos\biggl(\pi\frac{\phi}{\phi_0}\biggr) + \sum_{i=1}^N S_{i,2}\phi_{i,2}\sin\biggl(\pi\frac{\phi}{\phi_0}\biggr),
\end{split}
\end{equation}
where $C = 2e^2/E_C$ is the junction capacitance. The phase-related flux $\phi = \frac{\phi_0}{2\pi}\varphi$ (with $\phi_0=h/2e$ being the flux quantum) and the charge $Q$ on the junction electrodes are conjugate variables obeying $[\phi,Q]=i\hbar$. Two independent environmental baths are separately coupled to $\cos\biggl(\pi\frac{\phi}{\phi_0}\biggr)$ and $\sin\biggl(\pi\frac{\phi}{\phi_0}\biggr)$, with associated strengths $S_{i,1}$ and $S_{i,2}$. Upon integrating out these bosonic degrees of freedom, the effective action reduces to that of Eq.~(1).

The spectral function of the QP model is defined by $\frac{2\hbar^2}{\pi^2}J(\omega) = \sum_{i=1}^N S_{i,j}^2 \frac{\hbar}{2C_{i,j}\omega_{i,j}}\delta(\omega-\omega_{i,j})$, with bath frequencies $\omega_{i,j} = 1/\sqrt{L_{i,j}C_{i,j}}$ for $j=1,2$. Thus, it becomes evident that $\alpha$ measures the effective coupling between the system and the environmental degrees of freedom, i.e., a set of LC circuits. The cutoff frequency $\omega_D$ is associated with the highest energy scale in the problem, related to the Fermi energy of the two superconductors. 

\section{The worldline Monte Carlo (WLMC) method}
The proposed WLMC framework relies on the path-integral formulation. At a finite temperature $T = 1/(k_B\beta)$, the partition function for the action ${\cal S}_{QP}$ takes the form of a functional integral over $\beta\hbar$-periodic imaginary-time trajectories $\varphi(\tau)$:
\begin{equation}\label{eqn:partiton}
    Z_{QP} = \int\limits_{\varphi(0) = \varphi(\beta\hbar)} {\cal D} [\varphi(\tau)] \,e^{-\mathcal{S}_{QP}[\varphi(\tau)]/\hbar}.
\end{equation}
Consequently, the thermal expectation value of a physical observable $A$, expressed as a functional $A[\varphi(\tau)]$, is evaluated via the following:
\begin{equation}\label{eqn:mean_value}
    \langle A \rangle = \frac{1}{Z_{QP}}\int\limits_{\varphi(0) = \varphi(\beta\hbar)} {\cal D} [\varphi(\tau)] \, A[\varphi(\tau)]e^{-{\cal S}_{QP}[\varphi(\tau)]/\hbar}.
\end{equation}
To numerically access these path integrals, we map the quantum system onto a one-dimensional classical chain through Trotterization. The discretized action reads:
\begin{equation}\label{eqn:disc_action}\begin{split}
    {\cal S}_{QP}^{\text{dis}}[\{\varphi_m\}] = -E_J \Delta \tau \sum_{i=0}^{N-1}\cos\varphi_i+
    \sum_{i<j} K^{\text{dis}}_{ij}\sin^2\biggl(\frac{\varphi_i-\varphi_j}{4}\biggr)+\frac{\hbar^2}{4E_C \Delta \tau}\sum_{i=0}^{N-1}\bigl(\varphi_i - \varphi_{i+1}\bigr)^2,
    \end{split}
\end{equation}
where $\Delta \tau = \beta\hbar/N$ is the imaginary-time step, $N$ is the number of Trotter slices, and the modified kernel is $K^{\text{dis}}_{ij} = 2\Delta \tau^2 K\bigl(\Delta\tau (i-j)\bigr)$. The continuous trajectories $\varphi(\tau)$ are therefore replaced by discrete configurations $\{\varphi_m\}$ evaluated at $\tau_i = i\Delta \tau$, with periodic boundary conditions enforced by $\varphi_{N} = \varphi_0$. Consequently, the partition function simplifies to a discrete sum over all configuration sets:
\begin{equation}\label{eqn:disc_part}\begin{split}
    Z_{QP}^{\text{dis}} = \sum_{\{\varphi_m\}}e^{-{\cal S}_{QP}^{\text{dis}}[\{\varphi_m\}]/\hbar} 
    \end{split},
\end{equation}
and the expectation value becomes:
\begin{equation}\label{eqn:disc_mean}\begin{split}
    \langle A^{\text{dis}} \rangle =\frac{1}{Z_{QP}^{\text{dis}}} \sum_{\{\varphi_m\}}A(\{\varphi_m\})e^{-{\cal S}^{\text{dis}}_{QP}[\{\varphi_m\}]/\hbar} 
    \end{split}.
\end{equation}
This finite-difference approximation introduces a systematic Trotter error of $\mathcal{O}(\Delta \tau)$. In this work, we set the $\Delta \tau = 2\hbar/3E_C$. The original quantum problem is now rigorously mapped onto a 1D classical chain of interacting continuous phase variables. To efficiently compute \eqref{eqn:disc_mean}, our WLMC protocol constructs a Markov chain that samples the discretized worldlines $\{\varphi_m\}$ according to the Boltzmann weight $p(\{\varphi_m\}) \propto e^{-{\cal S}_{QP}^{\text{dis}}[\{\varphi_m\}]/\hbar}$.

To overcome critical slowing down, we employ a specialized cluster-update scheme inspired by the Wolff algorithm \cite{Wolff_PRL} and adapted for continuous variables by Werner and Troyer \cite{Troyer_MC,Troyer_PRL}. An update is initialized by choosing a random root slice $j$ and a reflection axis $a \in [-L\pi, L\pi]$. Sites added to the cluster undergo a parity reflection across this axis:
\begin{equation}\label{eqn:update}\begin{split}
    \varphi_l^\text{new} = 2a -\varphi_l^{\text{old}} 
    \end{split}.
\end{equation}
The cluster is expanded iteratively by establishing bonds between an unvisited site $u$ and a current cluster member $v$, using the Wolff-like acceptance probability:
\begin{equation}\label{eqn:prob_wolff}\begin{split}
    P_W(u|v)  = \max\biggl\{0, 1 -e^{-[{\cal S}_R(\varphi^{\text{old}}_u,\varphi_v^{\text{new}})-{\cal S}_R(\varphi^{\text{new}}_u,\varphi_v^{\text{new}})]/\hbar} \biggr\},
    \end{split}
\end{equation}
where the non-local and kinetic contributions are encapsulated within the retarded action:
\begin{equation}\label{eqn:ret_action}\begin{split}
    {\cal S}_R(\varphi_u,\varphi_v) =   K^{\text{dis}}_{uv} \sin^2 \biggl( \frac{\varphi_u-\varphi_v}{4} \biggr)  +  \frac{\hbar^2}{4E_C\Delta \tau} (\delta_{v,u+1} + \delta_{v,u-1}) (\varphi_u-\varphi_v)^2. 
    \end{split}
\end{equation}
Once the cluster is fully constructed, the global reflection is accepted or rejected based on the Metropolis criterion dictated by the static potential:
\begin{equation}\label{eqn:prob_metr}\begin{split}
    P_M\bigl(\{\varphi_m^{\text{new}}\} \to \{\varphi_m^{\text{old}}\}\bigl)  = \min\biggl\{1,  e^{-[{\cal S}_J(\{\varphi_m^{\text{new}}\})-{\cal S}_J(\{\varphi_m^{\text{old}}\})]/\hbar} \biggr\},
    \end{split}
\end{equation}
where ${\cal S}_J\bigl(\{\varphi_m\}\bigr) = -E_J\Delta\tau\sum_{i=0}^{N-1}\cos\varphi_i$ represents the local Josephson energy.

Our simulation alternates between two specific cluster micromoves. In the first, the reflection axis $a$ is constrained to the symmetry points of the Josephson potential, i.e., $a = k\pi$ with $k \in [-L, L]$ (optimally selecting among the three multiples of $\pi$ closest to the root). This targeted choice automatically yields a Metropolis acceptance probability of $1$ in \eqref{eqn:prob_metr} \cite{Troyer_MC, Troyer_PRL}. However, relying exclusively on symmetry-axis reflections breaks ergodicity. To restore it, we introduce a second micromove where $a$ is drawn uniformly at random near the root. Although this continuously-distributed axis can produce smaller, sometimes single-site clusters, it strictly guaranties that the Markov chain is ergodic. The boundary parameter $L$ is dynamically adjusted during thermalization to fully encompass the trajectory spread, effectively simulating the extended phase space without artificial cutoff effects.

A complete Monte Carlo sweep is defined as a sequence of these micromoves sufficient to attempt an update on the $\mathcal{O}(N)$ lattice sites. However, the presence of long-range interactions in ${\cal S}_R$ fundamentally scales the cluster construction time to $\mathcal{O}(N^2)$ per sweep, severely hindering low-temperature (large $\beta$) simulations. To systematically bypass this computational bottleneck, we integrate the Luijten-Bl\"{o}te (LB) algorithm \cite{Luijten1, Luijten2}. The core idea relies on the definition of a configuration-independent bounding function $F_{\left|u-v\right|}$ that satisfies the following:
\begin{equation}
    \begin{split}
        F_{\left|u-v\right|} \ge {\cal S}_R(\varphi_u^{\text{old}},\varphi_u^{\text{new}}) - {\cal S}_R(\varphi_u^{\text{new}},\varphi_u^{\text{new}}),
    \end{split}
\end{equation}
depending strictly on the lattice distance between $u$ and $v$. The cluster generation is then split into a two-tier process. Initially, we populate the lattice with provisional bonds drawn from the upper-bound probability distribution:
\begin{equation}\label{eqn:luijten_provv}\begin{split}
    P_{1LB}(u|v)  = 1 -e^{-F_{\left|u-v\right|}/\hbar}. 
    \end{split}
\end{equation}
Rather than evaluating this probability for all pairs, we sample the distance to the next provisional bond. Given a starting site $u$, the joint probability of skipping sites $u+1$ through $u+k-1$ and firmly placing a bond at $u+k$ is:
\begin{equation}\label{eqn:luijten_provv_up_to_k}\begin{split}
    P(k) = e^{-(F_1+F_2+\dots+F_{k-1})/\hbar}\biggl(1-e^{-F_{k}/\hbar}\biggr),
    \end{split}
\end{equation}
yielding the cumulative probability distribution:
\begin{equation}\label{eqn:luijten_cum}\begin{split}
    C(k) = \sum_{k'=1}^{k} P(k').
    \end{split}
\end{equation}
By sampling a uniform random number, we determine the first provisional bond at $u+k_1$ (with probability $C(k_1)-C(k_1-1)$), or stop bond formation entirely (with probability $1-C(N)$). If $u+k_1$ is successfully linked, the probability of the next consecutive bond at distance $k$ requires rescaling by the remaining probability space:
\begin{equation*}\begin{split}
    P_2(k) = e^{-(F_{k_1+1}+F_{k_1+2}+\dots+F_{k-1})/\hbar}\biggl(1-e^{-F_{k}/\hbar}\biggr) = \frac{P(k)}{1-C(k_1)},
    \end{split}
\end{equation*}
with its corresponding updated cumulative:
\begin{equation}\begin{split}
    C_2(k) = \sum_{k'=k_1+1}^{k} P(k') = \frac{C(k)-C(k_1)}{1-C(k_1)}.
    \end{split}
\end{equation}
This stochastic progression is repeated iteratively until the bond-proposing mechanism terminates. Crucially, because $F_{\left|u-v\right|}$ is purely a function of inter-site distance, the cumulative distributions can be precomputed. Utilizing a bisection search, the next bonded site is identified in $\mathcal{O}(\log N)$ operations rather than $\mathcal{O}(N)$.

Finally, in the second stage of the LB scheme, the proposed topological bonds are formally accepted and added to the cluster with the conditional probability:
\begin{equation*}
\begin{split}
    P_{2LB}(u|v)  =  \frac{\max \biggl\{0, 1-e^{-[{\cal S}_R(\varphi^{\text{old}}_u,\varphi_v^{\text{new}})-{\cal S}_R(\varphi^{\text{new}}_u,\varphi_v^{\text{new}})]/\hbar}\biggr\}}{1 -e^{-F_{\left|u-v\right|}/\hbar}}. 
    \end{split}
\end{equation*}

\section{Variational approach à la Feynman}
This section details a Feynman-type variational method, conceptually analogous to the framework developed for the Fr\"ohlich polaron \cite{Feynman_var}. We consider the total Hamiltonian given in Eq.~(\ref{eqn:hamiltonian}):
\begin{equation}\label{eqn:rep_H_supp}
H_{QP} = \frac{Q^2}{2C} - E_J\cos\biggl(2\pi\frac{\phi}{\phi_0}\biggr) + \sum_{j=1}^2 \sum_{i=1}^N \hbar \omega_{i,j} a^\dagger_{i,j} a_{i,j}
    + \sum_{i=1}^N \lambda_{i,1} (a_{i,1} + a^\dagger_{i,1}) \cos\biggl(2\pi\frac{\phi}{2\phi_0}\biggr) + \sum_{i=1}^N \lambda_{i,2} (a_{i,2} + a^\dagger_{i,2}) \sin\biggl(2\pi\frac{\phi}{2\phi_0}\biggr),
\end{equation}
with coupling constants defined as $\lambda_{i,j} = S_{i,j} \sqrt{\frac{\hbar}{2C_{i,j}\omega_{i,j}}}$. For clarity, we partition the system into the bare Cooper pair box Hamiltonian $H_{CPB} = Q^2/2C -E_J \cos(2\pi \phi/\phi_0)$ and the non-interacting baths $H_{B_j} = \sum_{i=1}^N \hbar \omega_{i,j} a^\dagger_{i,j} a_{i,j}$. Moreover, $H_{I_1} = \sum_{i=1}^N \lambda_{i,1} (a_{i,1} + a^\dagger_{i,1}) \cos\biggl(2\pi\frac{\phi}{2\phi_0}\biggr)$ and $H_{I_2} = \sum_{i=1}^N \lambda_{i,2} (a_{i,2} + a^\dagger_{i,2}) \sin\biggl(2\pi\frac{\phi}{2\phi_0}\biggr)$ are the respective interaction components. 

Following ordered operator calculus \cite{Feynman1_var}—the algebraic counterpart to the path integral formalism—we first evaluate the full partition function \cite{Fetter, Mahan}:
\begin{equation}
Z_{QP}=\Tr \left[ e^{-\beta H}\right] =Z_{0} U(\beta),
\end{equation}
where the evolution operator is $U(\beta)=\langle T_{\tau} e^{-\int_{0}^{\beta} d\tau^{\prime}H_{I}^{(0)}(\tau^{\prime})}\rangle_{0}$. Here, $Z_{0}$ represents the partition function of the uncoupled system ($H_{0}=H_{CPB}+H_{B_1} +H_{B_2}$), $T_{\tau}$ denotes imaginary-time ordering, $H_{I}^{(0)}$ is the total interaction $H_I = H_{I_1} + H_{I_2}$ transformed into the interaction picture, and $\langle ... \rangle_{0}$ indicates the thermal average over $H_{0}$. Because the unperturbed basis factorizes, the bosonic bath degrees of freedom can be traced out exactly via the Bloch-DeDominicis theorem \cite{Fetter, Mahan}, yielding:
\begin{equation}\label{eq:zeta}
  Z_{QP}=Z_{CPB}Z_{B_1} Z_{B_2}\langle T_{\tau} e^{\Phi}\rangle_{CPB},
\end{equation}
where the effective action term is:
\begin{equation}\begin{split}\label{eq:phi}
  \Phi=&\frac{1}{2}\int_0^{\beta\hbar }d\tau \int_0^{\beta\hbar} d\tau^{\prime} \cos \biggl( \frac{2\pi \phi^{(0)}(\tau)}{2\phi_0} \biggr)K(\tau-\tau^{\prime}) \cos \biggl( \frac{2\pi \phi^{(0)}(\tau^{\prime})}{2\phi_0} \biggr) \\
  +& \frac{1}{2}\int_0^{\beta\hbar }d\tau \int_0^{\beta\hbar} d\tau^{\prime} \sin \biggl( \frac{2\pi \phi^{(0)}(\tau)}{2\phi_0} \biggr)K(\tau-\tau^{\prime}) \sin \biggl( \frac{2\pi \phi^{(0)}(\tau^{\prime})}{2\phi_0} \biggr).
\end{split}
\end{equation}
The explicit form of $K(\tau)$ is detailed in the main text, derived by taking the continuum limit $N\to \infty$ and employing Eqs.~(2)-(4). Crucially, the expectation value in Eq.~(\ref{eq:zeta}) is now evaluated solely over the $H_{CPB}$ subspace. Furthermore, unlike the standard path-integral formulation of Eq.~(1), the argument in Eq.~(\ref{eq:phi}) explicitly retains the quantum phase operators rather than replacing them with classical variables. Up to this point, the treatment is exact.

Although a direct perturbative expansion of the exponential is possible, the proliferation of Feynman diagrams at strong coupling makes the approach intractable. To avoid this limitation, one can proceed in two complementary ways. The first route consists of invoking Feynman’s variational principle, introducing a suitably chosen trial Hamiltonian, $H_{tr}$, which captures the dominant physical processes while remaining analytically manageable. The true macroscopic bath is replaced by a truncated set of $N$ fictitious bosonic modes:
\begin{equation}\label{eq:definitionHtrial}
  H_{tr}=H_{CPB} + \sum_{j=1}^2\sum_{i=1}^{N}\hbar \tilde{\omega}_{i} a_{i,j}^{\dagger}a_{i,j}+ \cos \biggl( \frac{2\pi \phi} {2\phi_0} \biggr)\sum_{i=1}^N \tilde{\lambda}_{i}\left(a_{i,1}^\dagger+a_{i,1}\right) + \sin \biggl( \frac{2\pi \phi} {2\phi_0} \biggr)\sum_{i=1}^N \tilde{\lambda}_{i}\left(a_{i,2}^\dagger+a_{i,2}\right).
\end{equation}
The effective frequencies $\tilde{\omega}_{i}$ and the couplings $\tilde{\lambda}_{i}$ can be chosen in a variational way. Indeed, since in $H_{tr}$ the trial interactions remain linear in the bosonic operators, these fictitious modes can be integrated exactly as before, resulting in:
\begin{equation}\label{eq:zetatrial}
  Z_{tr}=Z_{CPB} Z_{{B_1, tr}} Z_{{B_2, tr}} \langle T_{\tau} e^{\Phi_{tr}}\rangle_{CPB},
\end{equation}
where
\begin{equation}\begin{split}\label{eq:phitrial}
  \Phi_{tr}=&\frac{1}{2}\int_0^{\beta\hbar }d\tau \int_0^{\beta\hbar} d\tau^{\prime} \cos \biggl( \frac{2\pi \phi^{(0)}(\tau)}{2\phi_0} \biggr)K_{tr}(\tau-\tau^{\prime}) \cos \biggl( \frac{2\pi \phi^{(0)}(\tau^{\prime})}{2\phi_0} \biggr) \\
  +& \frac{1}{2}\int_0^{\beta\hbar }d\tau \int_0^{\beta\hbar} d\tau^{\prime} \sin \biggl( \frac{2\pi \phi^{(0)}(\tau)}{2\phi_0} \biggr)K_{tr}(\tau-\tau^{\prime}) \sin \biggl( \frac{2\pi \phi^{(0)}(\tau^{\prime})}{2\phi_0} \biggr).
\end{split}
\end{equation}
Here, the kernel $K_{tr}(\tau)=\sum_{i=1}^N \tilde{\lambda}_i^2 \frac{ \cosh \left[ \tilde{\omega}_i \left( \frac{\beta}{2}-\tau \right)\right]}{\sinh \left( \frac{\beta \tilde{\omega}_i}{2} \right)}$ encodes the propagator and interaction strengths of the trial modes. We emphasize that $\Phi_{tr}$ and $\Phi$ share the same formal structure; they differ only in the specific form of the kernel that enters their definition.

Next, we define the generating function $f(x)=-T \log \langle T_{\tau}e^{\Phi_{tr}+x \left(\Phi-\Phi_{tr} \right)}\rangle_{CPB}$. It is straightforward to demonstrate that its second derivative with respect to $x$ is strictly negative in the interval $x \in [0,1]$ \cite{bogoliubov_var}. This inherent concavity ensures the inequality $f(1)-f(0) \le f^{\prime}(0)$, which establishes a rigorous upper bound for the true free energy $F = -T \log Z_{QP}$: 
\begin{equation}\label{eq:freeenergy}
  F-F_{B_1} - F_{B_2} \le F_{tr}-F_{{B_1, tr}} - F_{{B_2, tr}} -T \frac{ \langle T_{\tau} e^{\Phi_{tr}} \left(\Phi-\Phi_{tr}  \right)\rangle_{CPB}}{ \langle T_{\tau} e^{\Phi_{tr}}\rangle_{CPB}}.
\end{equation}
This bound is mathematically equivalent to the Feynman-Jensen inequality originally derived for the polaron problem \cite{Feynman_var,Feynman2_var} and acts as a direct generalization of the standard Bogoliubov bound \cite{Feynman2_var}. The optimal trial Hamiltonian is then determined by minimizing the variational free‑energy functional, thereby providing a non-‑perturbative approximation that effectively resums an infinite subset of diagrams and yields controlled estimates even deep in the strong‑coupling regime.

We emphasize that, although the trial kernel $K_{tr}(\tau)$ decays exponentially at large $\tau$ for any finite value of $N$, the physical kernel $K(\tau)$ instead exhibits a power‑law tail. This qualitative difference is crucial: the exponential decay reflects the finite‑gap structure imposed by the variational ansatz, whereas the true kernel retains long‑range temporal correlations that cannot be captured by any finite‑$N$ truncation. It is then clear that, as the temperature is lowered, the number of fictitious modes must be increased accordingly. Since the variational principle requires the exact diagonalization of the trial Hamiltonian, followed by a minimization over the frequencies and couplings of the fictitious baths, it is evident that the procedure becomes increasingly demanding as $N$ grows. 

The second route avoids these bottlenecks by recognizing that the system’s physics is primarily governed by the long‑range behavior of the kernel $K(\tau)$. Then the parameters $\{\tilde{\omega}_{i}, \tilde{\lambda}_{i}\}$ can be determined by fitting the trial kernel $K_{tr}(\tau)$ directly to the exact kernel $K(\tau)$ over the interval $[\tau_0, \beta/2]$, i.e., by focusing the attention only on the power-law decay bahavior of the kernel $K(\tau)$. For the numerical results presented in this work, we restricted the expansion to $N = 2$ modes per bath and set the lower cutoff to $\tau_0 E_C/\hbar = 3$. This choice allowed us to obtain a successful agreement between the spectra of $H_{tr}$ and $H_{QP}$ within the linear response regime at $E_C \beta=100$, as shown in the main text.

\section{Density-matrix renormalization group algorithm}
We compute the effective energy bandwidth for different values of the coupling $\alpha$ by obtaining the ground-state energy of the full system, including the degrees of freedom of the two baths, using the density-matrix renormalization group (DMRG) algorithm.

The DMRG algorithm \cite{white1992density,schollwock2005density,schollwock2011density} is an adaptive variational method for optimizing a matrix product state (MPS) representation of the low-energy eigenstates of a large Hermitian Hamiltonian $H$. In the two-site version of the algorithm, two neighboring MPS tensors are iteratively combined into a single two-site tensor and optimized. The local optimization is typically performed using iterative eigensolvers such as the Lanczos or Davidson algorithms. The optimized tensor is then factorized using either a singular value decomposition or a density-matrix decomposition. This restores the MPS structure while allowing the bond dimension to be adapted during the factorization step, with negligible singular values being truncated according to a prescribed accuracy threshold. For this purpose, we recast $H_{QP}$ in a different representation. Specifically, to make DMRG more efficient, the Hamiltonian must be formulated in a site representation and decomposed into a sum of terms acting on at most two sites. 

Using the $2\phi_0$ periodicity of $H_{QP}$, the states of the system can be expanded on a plane-wave basis as
\begin{equation}
    \langle\phi|\psi_{u,k} \rangle = \frac{e^{i\phi(k+\pi u/\phi_0 )}}{\sqrt{2M\phi_0}},
\end{equation}
where $M$ is a large, odd integer defining the number of unit cells, each of them with length $2\phi_0$. The index $u$ is an integer, and $k = \frac{2\pi}{2\phi_0} \frac{m}{M}$ represents the quasicharge, with $m \in [-(M-1)/2, (M-1)/2]$.

The Hamiltonian in Eq.~\eqref{eqn:rep_H_supp} can be further projected onto the eigenbasis of the unitary translation operator $U = e^{i\frac{\varphi}{2}}$, where $\varphi=\frac{2 \pi \phi}{\phi_0}$. This operator does not mix states with different quasicharges $k$, therefore, we can solve the eigenvalue problem for each $k$-sector:
\begin{equation}
    U |V_{n,k} \rangle = \lambda_{n} |V_{n,k} \rangle. 
\end{equation}

The unitarity of $U$ ensures that its eigenvalues take the form $\lambda_n = e^{iS_n}$, with $S_n$ being purely real. Introducing the creation and annihilation operators, $d^{\dagger}_{n,k}$ and $d_{n,k}$, associated with the state $|V_{n,k}\rangle$, the trigonometric potential terms can be expressed in second quantization as:
\begin{equation*}\begin{split}
    &\cos\biggl(2\pi \frac{\phi}{2\phi_0} \biggr) = \sum_{n,k} d^{\dagger}_{n,k} d_{n,k} \cos(S_n),\\
    &\sin\biggl(2\pi \frac{\phi}{2\phi_0} \biggr) = \sum_{n,k} d^{\dagger}_{n,k} d_{n,k} \sin(S_n),\\
    &\cos\biggl(2\pi \frac{\phi}{\phi_0} \biggr) = \sum_{n,k} d^{\dagger}_{n,k} d_{n,k} \cos(2 S_n).
\end{split}
\end{equation*}
Furthermore, since the charge operator $Q$ is diagonal in the plane-wave representation, it preserves the quasicharge $k$. The kinetic energy of the phase particle is represented by:
\begin{equation*}\begin{split}
    \frac{Q^2}{2C} = \sum_{k,n,n'} d^{\dagger}_{n,k} d_{n',k} \biggl \langle  V_{n,k} \biggl| \frac{Q^2}{2C} \biggr|V_{n',k} \biggl \rangle .
\end{split}
\end{equation*}
As a consequence, the full Hamiltonian of the system can be recast as follows:
\begin{equation}\label{eqn:Holstein}
    \begin{split}
        H_{QP} = &-E_J \sum_{n,k} d^{\dagger}_{n,k} d_{n,k} \cos(2 S_n) + \sum_{k,n,n'} d^{\dagger}_{n,k} d_{n',k} \biggl \langle  V_{n,k} \biggl| \frac{Q^2}{2C} \biggr|V_{n',k} \biggl \rangle + \sum_{j=1}^2 \sum_{i=1}^N \hbar \omega_{i,j} a^\dagger_{i,j} a_{i,j}\\
    &+ \sum_{i=1}^N \lambda_{i,1} (a_{i,1} + a^\dagger_{i,1}) \sum_{n,k} d^{\dagger}_{n,k} d_{n,k} \cos(S_n) + \sum_{i=1}^N \lambda_{i,2} (a_{i,2} + a^\dagger_{i,2}) \sum_{n,k} d^{\dagger}_{n,k} d_{n,k} \sin(S_n).
    \end{split}
\end{equation}
The Hamiltonian derived in Eq.~\eqref{eqn:Holstein} describes, for any fixed value of $k$, a particle moving in a lattice (the sites are labeled by the index $n$) in a potential and coupled to two Holstein-like (diagonal in the position operator of the particle on each site) phonon branches represented by the bosonic fields $a_{i,1}$ and $a_{i,2}$.

For each fixed value of the quasicharge $k$, the Hamiltonian in Eq.~\eqref{eqn:Holstein} defines a corresponding Hamiltonian $H_{QP}(k)$. We represent this Hamiltonian as a matrix product operator and optimize the associated ground state in MPS form. The MPS chain is organized as follows: the leftmost $N$ sites correspond to the modes of phonon branch 1, the following $L$ sites represent the one-dimensional lattice on which the particle moves, and the rightmost $N$ sites correspond to the modes of phonon branch 2. We adopt a star geometry, in which each bosonic mode couples directly to the particle density on every lattice site. This produces long-range interactions along the MPS chain.

Our DMRG calculations are performed using the ITensor Library \cite{fishman2022itensor}. In the implementation, we control several accuracy parameters, including the maximum and minimum bond dimensions of the MPS, the truncation cutoff used during singular value decompositions or density-matrix diagonalizations, the maximum number of Davidson iterations in each local optimization step, and the magnitude of the noise term added to the density matrix to improve convergence. We verified that the final ground-state energy obtained after the DMRG sweeps is converged to at least 9 significant digits (more than single precision). This level of accuracy is maintained also in the vicinity of the transition, where convergence is slower.

In numerical calculations, we use $N=200$ bosonic modes for each phonon branch and up to $L=121$ lattice sites, labeled by the index $n$ in Eq.~\eqref{eqn:Holstein}. The spectral densities of the two baths are Ohmic ($s=1$). The bath frequencies are discretized with a linear grid refined at low frequencies, which improves the convergence of the stationary state. The local Hilbert space of each bosonic mode is truncated to $N_{\mathrm{ph}}=3$ Fock states.

For each quasicharge $k$, the Hamiltonian $H_{QP}(k)$ is constructed by fixing the $k$-dependent coefficients $\cos(S_n)$, $\sin(S_n)$, $\cos(2S_n)$, and
$
\left\langle V_{n,k} \left| \frac{Q^2}{2C} \right| V_{n',k} \right\rangle
$
appearing in Eq.~\eqref{eqn:Holstein}. We then use DMRG to compute the ground-state energy $E_0(k)$ of $H_{QP}(k)$. The effective bandwidth is defined as
$
\Delta E = E_0(k_+) - E_0(k_-),
$
with $\hbar k_+=e/2$ and $\hbar k_-=0$, as shown in the inset of Fig.~1(b) of the main text.

\section{Linear response of a polaronic model}
In this section, we outline the main differences in the linear response between the model presented in the main text and an exactly solvable polaronic model. To this end, we introduce the Hamiltonian
\begin{equation}
    H = \frac{p^2}{2m} + \sum_{i = 1}^N \biggl(\frac{p_i^2}{2m_i} + \frac{k_i}{2} (x-x_i)^2\biggr),
\end{equation}
which describes an electron of mass $m$ and charge $e$ (the system) interacting with a large bath ($N\gg1$) of harmonic oscillators (the environment). The bath represents the crystal lattice; consequently, the linear coupling term captures the phonon-mediated interaction between the electron and lattice distortions at the lowest order. This polaronic model~\cite{Weiss2021} is widely known as the Caldeira-Leggett model (CL) ~\cite{PhysRevLettCL1}. Assuming an Ohmic environment, the spectral density is given by
\begin{equation}
    \sum_{i=1}^N \frac{k_i^2}{2m_i\omega_i} \delta(\omega - \omega_i) = m\alpha\omega_c \omega \theta(\omega_D-\omega),
\end{equation}
where $\omega_i^2 = k_i/m_i$ and $\omega_D$ is the Debye frequency. Here, the dimensionless quantity $\alpha$ acts as the electron-phonon coupling constant. It plays a role analogous to the homonymous parameter in the main text, quantifying the system-environment interaction strength.

To formulate the linear response theory for conductivity under an electric field bias, we introduce the current operator $I = ep/m$. Here, $x$ is the electron displacement, $e x$ is the dipole momentum, and $p$ is the conjugate momentum to $x$, such that $[x, p] = i\hbar$. Following the convention of Eq.~(13) in the main text, we set $B = ex$ and $A = dB/dt = I$. The response function is then
\begin{equation}\label{eqn:kubo_supp}
    \Psi_I (z) = \frac{i}{z} \biggl[ \Pi_I(z) + \frac{e^2}{m} \biggr],
\end{equation}
where we used $\langle k \rangle = \frac{i}{\hbar}[I, ex] = e^2/m$. Eq.~\eqref{eqn:kubo_supp} is the Kubo formula for the optical conductivity $\Psi_I(z)$. As discussed in the main text, taking the limit $z = \omega + i0^+$ yields
\begin{equation}
    \mathrm{Re}[\Psi_I(\omega)] = \pi D_I\delta (\omega) + \Psi_{I,\mathrm{reg}} (\omega). 
\end{equation}
Furthermore, via the Lehmann representation, it can be shown that the regular part must satisfy the following sum rule:
\begin{equation}\label{eqn:sum_rule_supp}
    \int_0^{\infty} d\omega \, \Psi_{I, \mathrm{reg}} (\omega) = \frac{\pi}{2}\biggl[\frac{e^2}{m} - D_I \biggr].
\end{equation}

We now distinguish two cases. First, we consider the decoupled limit ($\alpha = 0$), where the model is analytically solvable. We start from the plane-wave eigenstates $|n\rangle$ of a system of size $L$ with periodic boundary conditions, $\langle x|n\rangle = e^{ik_nx}/\sqrt{L}$, where $k_n = 2\pi n/L$ and $n \in \mathbb{Z}$. In the thermodynamic limit ($L \to \infty$), one can show that $\Pi_I(i \omega_n \neq 0) = 0$ and $\Pi_I(i\omega_n = 0) = e^2/m$. Consequently, $D_{I,M} = 0$ and $D_I = e^2/m$. Because $\Psi_{I,\mathrm{reg}}(\omega) \ge 0$ and the sum rule~\eqref{eqn:sum_rule_supp} must be fulfilled, the regular contribution to the conductivity vanishes for all $\omega$, yielding
\begin{equation}
    \mathrm{Re}[\Psi_I(\omega)] = \pi D_I \delta(\omega).
\end{equation}

The same result can be obtained by explicitly evaluating the regular part of the optical conductivity, using the plane-wave basis. 

When the environment is coupled ($\alpha \neq 0$), the Hamiltonian can be exactly diagonalized. Indeed, it turns out to be described as the sum of $N+1$ independent harmonic oscillators (normal modes) with frequencies $\tilde \omega_j$, whose values are given by the roots of the equation
\begin{equation}
    \tilde \omega_j^2 - \sum_{i=1}^N\frac{m_i}{m} \biggl(\frac{\omega_i^2 \tilde \omega_j^2}{\tilde \omega_j^2-\omega_i^2}\biggr) = 0,
\end{equation}
for $j=1, \dots, N+1$. We  note that a zero-frequency solution, $\tilde \omega_1 = 0$, is always present. Furthermore, the elements of the first row of the orthogonal transformation matrix $P_{i,j}$ satisfy
\begin{equation} \label{eqn:p1i}
\begin{split}
    &P^2_{1,j} = \frac{1}{1+\sum_{i=1}^N \frac{m_i}{m}\frac{\omega_i^4}{\bigl(\tilde \omega_j^2-\omega_i^2\bigr)^2}},\\
    &\lim_{N \to \infty} P^2_{1,1} = 0. 
\end{split}
\end{equation}

Exploiting the Lehmann representation in the large-$N$ limit, we obtain, at any finite temperature, the following result:
\begin{equation}\label{eqn:psireg_supp}
    \Psi_{I, \mathrm{reg}}(\omega) = \frac{\pi e^2}{2m}\sum_{i=2}^{N+1} \bigl[\delta(\omega-\tilde \omega_i) + \delta(\omega+\tilde \omega_i)\bigr] P_{1,i}^2.
\end{equation}

 We emphasize that the summation starts from $i=2$, as required by Eq.~\ref{eqn:p1i}.

\begin{figure}
\centering
\includegraphics[]{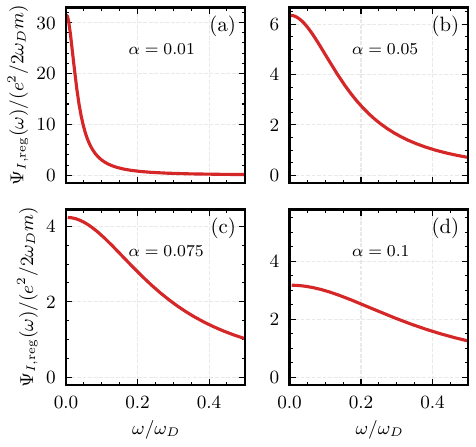}
\caption{Regular contribution to the conductivity at any finite temperature, $\Psi_{I,\mathrm{reg}}(\omega)$, for various values of the coupling $\alpha$. For finite coupling ($\alpha \neq 0$), the response exhibits a Drude-like peak that progressively narrows as $\alpha \to 0$, ultimately collapsing into a pure Dirac delta distribution in the decoupled limit ($\alpha = 0$).
}\label{Fig:pol_psi}
\end{figure}
Fig.~\ref{Fig:pol_psi} shows the regular part of the optical conductivity for different values of $\alpha$. The plots reveal a Drude-like behavior of the regular contribution $\Psi_{I,\mathrm{reg}}(\omega)$. By substituting Eq.~\ref{eqn:psireg_supp} into the sum rule, we find $D_I = D_{I,M} = 0$. Consequently, $\Pi_I(i\omega_n \to 0) = -e^2/m$, making it a discontinuous function of $\alpha$ for $\alpha = 0$. Indeed,
\[
    -\frac{e^2}{m} = \Pi_I(i\omega_n \to 0, \alpha \ne 0) \neq \Pi_I(i\omega_n \to 0, \alpha = 0) = 0.
\]

\begin{figure}
\centering
\includegraphics[]{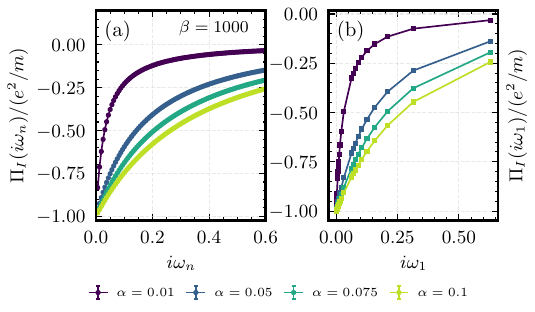}
\caption{(a) Normalized current-current correlation function $\Pi_I(i\omega_n)/(e^2/m)$ (with $e^2/m = \langle k\rangle$) versus Matsubara frequency $i\omega_n$ at inverse temperature $\beta \hbar \omega_D= 1000$ for various $\alpha$. (b) The same normalized function evaluated at the first Matsubara frequency, $i\omega_1 = 2\pi/\beta$ (in units of $\omega_D$), versus $i\omega_1$ for different $\alpha$. The comparison with panel (b) of Fig.~3 in the main text highlights the differences between the polaronic scenario and the framework that emerges in a small‑capacitance Josephson junction.
}\label{Fig:pol_pi}
\end{figure}
This highlights a major difference with respect to the model presented in the main text: for the polaronic model, $\Pi_I(i\omega_n \to 0)$ and $\Pi_I(i\omega_1)$ ($\Pi$ evaluated at the first Matusbara frequency, $\omega_1$) are discontinuous at $\alpha = 0$ (see Fig.~\ref{Fig:pol_pi}), while they remain continuous for the shunted junction. Our results definitely demonstrate that the CL model fails to capture the correct analytic structure of the Matsubara response

\section{Charge-Charge correlation function and Drude weight}
This appendix details the procedures used to evaluate the charge–charge correlation function $\Pi_Q(i\omega_n)$ and the Drude weight $D_Q$, both at finite and zero temperature. 

\subsection{Finite-Temperature Formalism}
We begin with the relation~\cite{Mahan}
\begin{equation}\begin{split}
    -&\int_0^{\beta \hbar} d\tau \, e^{i\omega_n \tau} \langle A(\tau) A(0) \rangle \\ = & \frac{1}{(i\omega_n)^2} \biggl\langle \biggl[A, \frac{\partial A}{\partial\tau} \bigg|_{\tau = 0} \biggr] \biggr\rangle + \frac{1}{(i\omega_n)^2} \int_0^{\beta \hbar} d\tau \, \biggl\langle \frac{\partial A}{\partial \tau} \frac{\partial A}{\partial \tau} \bigg|_{\tau = 0} \biggr\rangle,
\end{split}
\end{equation}
where $i\omega_n = 2n/\beta \hbar$ is the $n$-th Matsubara frequency ($n\neq 0$) and $A$ is a generic operator. Substituting $A = \phi$ into this expression yields
\begin{equation}\begin{split}
    -&\frac{1}{\hbar} \int_0^{\beta \hbar} d\tau \, e^{i\omega_n \tau }\langle\phi(\tau) \phi(0) \rangle \frac{(2\pi)^2}{\phi_0^2} \\ = & \frac{2 E_C}{ \hbar (i\omega_n)^2} - \frac{1}{(i\omega_n)^2} \frac{1}{\hbar^2} E_C^2 \int_0^{\beta \hbar} d\tau \, e^{i\omega_n\tau} \frac{1}{\hbar} \frac{\langle Q(\tau) Q(0)\rangle}{e^2},
\end{split}
\end{equation}
which can be recast as
\begin{equation}\label{eq:rel_mats_supp}
    \Pi_{\varphi}(i\omega_n) = \frac{2 E_C}{ \hbar (i\omega_n)^2} + \frac{1}{(i\omega_n)^2} \frac{1}{\hbar^2} \frac{E_C^2}{e^2} \Pi_{Q}(i\omega_n).
\end{equation}
Consequently, by computing $\Pi_{\varphi}(\tau)$ through WLMC and performing a Fourier transform, we can directly extract $\Pi_Q(i\omega_n)$ using Eq.~\eqref{eq:rel_mats_supp}.

To determine the Drude weight $D_Q$ at finite temperature, we adopt the approach introduced in~\cite{Evertz_PRB}. This involves fitting $\Pi_Q(i\omega_n)$ with a finite sum of Lorentzian functions to robustly extrapolate the zero-frequency limit, $\Pi_Q(i \omega_n \to 0)$. Finally, $D_Q$ is evaluated by substituting this extrapolated limit into Eq.~(9) of the main text. 

\subsection{Zero-Temperature Formalism}
At zero temperature ($T=0$), we instead rely on the method proposed by Kohn~\cite{Kohn_Drude}, which quantifies how strongly a quantum many‑body system responds to an infinitesimal twist in its boundary conditions, introduced through the parameter $Q_0$: 
\begin{equation}
    H_{Q_0} = \frac{\bigl(Q+Q_0e\bigr)^2}{2C} - E_J \cos \biggl(\frac{2\pi \phi}{\phi_0} \biggr) + \sum_{i=1}^2\bigl(H_{B_i} + H_{I_i}\bigr).
\end{equation}

In particular, the contribution $\frac{(Q+Q_0e)^2}{2C}$ provides: $\frac{ Q^2}{2C} + \frac{E_C Q_0^2}{4} + \frac{E_C Q_0}{2} \frac{Q}{e}$.

The energy of the ground state, i.e., $E_{\mathrm{GS}}$, becomes a function of $Q_0$. The Drude weight is then defined as the curvature of this energy at zero twist:
\begin{equation}
    \frac{D_Q}{2e^2/E_C} = \frac{2}{E_C} \frac{\partial^2 E_{\mathrm{GS}}}{\partial Q_0^2} \bigg|_{Q_0=0}.
\end{equation}
Using the Hellmann-Feynman theorem, this expression simplifies to
\begin{equation}
    \frac{D_Q}{e^2/E_C} = 2 \left[ 1 + \frac{\partial}{\partial Q_0} \left\langle \psi_{\mathrm{GS}, Q_0} \left| \frac{Q}{e} \right| \psi_{\mathrm{GS}, Q_0} \right\rangle \right]_{Q_0=0}, 
\end{equation}
where $|\psi_{\mathrm{GS},Q_0}\rangle$ represents the ground state of $H_{Q_0}$. ($|\psi_{\mathrm{GS},Q_0}\rangle$ and $\left\langle \psi_{\mathrm{GS}, Q_0} \left| \frac{Q}{e} \right| \psi_{\mathrm{GS}, Q_0} \right\rangle$ are evaluated using the DMRG approach in the subspace $k=0$ with periodic boundary conditions). 

The key quantity is $\frac{\partial}{\partial Q_0} \left\langle \psi_{\mathrm{GS}, Q_0} \left| \frac{Q}{e} \right| \psi_{\mathrm{GS}, Q_0} \right\rangle_{Q_0=0}$. In fact, $D_Q$ vanishes as this derivative approaches $-1$.

In the main text, we proved that the phase‑particle Hamiltonian can be mapped onto an effective spin‑boson model. This naturally raises the question of what quantity in the spin‑boson model plays the role of $D_Q$. To this aim, let us consider the lowest energy subspace, i.e. $k=0$, and the corresponding two level model:  
\begin{equation}
            H_{SB} = -\frac{\Delta}{2} \sigma_z + \sum_{i=1}^N \hbar \omega_i a_i^{\dagger}a_i + \sum_{i=1}^N g_{i} (a_i+a_i^{\dagger}) \sigma_x,
\end{equation}
with 
\begin{equation}
            \sum_{i=1}^N g_i^2 \delta(\omega-\omega_i)=\frac{\alpha}{2 \alpha_{c,j} } \omega^s \omega_C^{1-s} \theta(\omega_C-\omega). 
\end{equation}

Here, $\alpha_{c,j}$ denotes the critical coupling in the phase‑particle model (which depends on the ratio $\frac{E_j}{E_C}$), $\Delta$ is the effective gap of the spin, i.e., the two effective levels of the Cooper pair box, and $\omega_C$ is a cutoff energy. With this choice, the two level model also undergoes a QPT at $\alpha=\alpha_{c,j}$. The next step is to understand what corresponds, in the spin‑boson description, to the terms that depend on $Q_0$ in the phase‑particle Hamiltonian. To this end, we notice that the matrix elements of the  momentum operator of the phase particle, $Q/e$, fulfill the following properties: $\langle \psi_{n,k=0}| Q | \psi_{n,k=0} \rangle =0$, while $\langle \psi_{n,k=0}| Q | \psi_{m,k=0} \rangle \ne 0$ if $n\ne m$. Here, $| \psi_{n,k=0} \rangle$ denote the Bloch waves in the Cooper pair box model. Consequently, in the spin‑boson representation, the operator $\frac{Q}{e}$ maps to $\sigma_x$. Then, introducing the twist in the original Hamiltonian corresponds to applying a magnetic field along the $x$ direction in the mapped Hamiltonian: 
\begin{equation}
            H_{B_0} = H_{SB} + \frac{B_0^2 E_0^2}{4} + \frac{B_0 E_0}{2} \sigma_x,
\end{equation}
where $E_0$ is the spin gap in the absence of coupling with the environment, i.e., $E_0=\Delta$. The final step is to determine the counterpart, within the spin‑boson model, of the quantity $\frac{2}{E_C} \frac{\partial^2 E_{\mathrm{GS}}}{\partial Q_0^2} \bigg|_{Q_0=0}$. Using the Hellmann-Feynman theorem, we obtain: 
\begin{equation}
\frac{2}{E_0} \frac{\partial^2 E_{\mathrm{GS}}}{\partial B_0^2} \bigg|_{B_0=0}= \left[ 1 + \frac{\partial}{\partial B_0} \left\langle \psi_{\mathrm{GS}, B_0} \left| \sigma_x \right| \psi_{\mathrm{GS}, B_0} \right\rangle \right]_{B_0=0}.
\end{equation}

\begin{figure}
\centering
\includegraphics[]{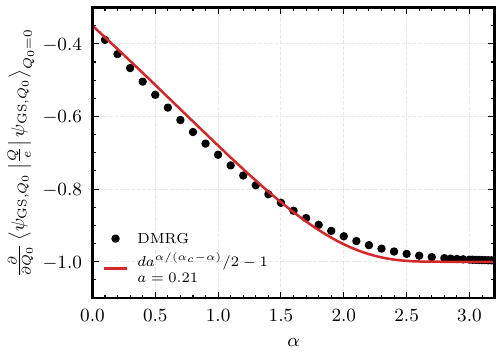}
\caption{DMRG evaluation of $\frac{\partial}{\partial Q_0} \left\langle \psi_{\mathrm{GS}, Q_0} \left| \frac{Q}{e} \right| \psi_{\mathrm{GS}, Q_0} \right\rangle_{Q_0=0}$ as a function of $\alpha$ (black dots). The solid curve is a fit to $d a^{\alpha/(\alpha_c-\alpha)}/2 - 1$ based on Eq.~\eqref{eqn:spin_boson_dq}, with fixed parameters $\alpha_c = 3.2$ and $d = 1.3$, yielding $a = 0.21$.
}\label{Fig:sb_fit}
\end{figure}

It is clear that this quantity represents $\frac{\Delta_{eff}}{\Delta}$, that is, the ratio between the effective gap of the two-‐level system and the bare spin gap. It goes from $1$, at $\alpha=0$, to $0$ at the onset of the QPT, where the quantity $\frac{\partial}{\partial B_0} \left\langle \psi_{\mathrm{GS}, B_0} \left| \sigma_x \right| \psi_{\mathrm{GS}, B_0} \right\rangle_{B_0=0}$ becomes $-1$. This condition reflects the fact that the spin behaves classically, i.e., quantum fluctuations are suppressed. On the other hand, it is well known \cite{Weiss1999} that the ratio $\frac{\Delta_{eff}}{\Delta}$ scales as $a^{\alpha/(\alpha_c-\alpha)}$, where $a$ is a parameter smaller than $1$, determined by the bare gap and the cutoff energy. Building on this mapping, it is natural to anticipate the following behavior for the quantity $\frac{\partial}{\partial Q_0} \left\langle \psi_{\mathrm{GS}, Q_0} \left| \frac{Q}{e} \right| \psi_{\mathrm{GS}, Q_0} \right\rangle_{Q_0=0}$: 
\begin{equation}\label{eqn:spin_boson_dq}
\frac{\partial}{\partial Q_0} \left\langle \psi_{\mathrm{GS}, Q_0} \left| \frac{Q}{e} \right| \psi_{\mathrm{GS}, Q_0} \right\rangle_{Q_0=0}=d a^{\alpha/(\alpha_c-\alpha)}/2 - 1,
\end{equation}
where $d$ is fixed by requiring that $\frac{D_Q}{e^2/E_C}$ matches the value obtained through exact diagonalization at $\alpha=0$. In Fig.~\ref{Fig:sb_fit} we plot $\frac{\partial}{\partial Q_0} \left\langle \psi_{\mathrm{GS}, Q_0} \left| \frac{Q}{e} \right| \psi_{\mathrm{GS}, Q_0} \right\rangle_{Q_0=0}$ together with the expected fit, which involves only a single free parameter, i.e. $a$. We emphasize that $\alpha_c$ is not a fitting parameter, but is instead fixed to the value obtained from the Monte Carlo analysis presented in the main text. The plot clearly illustrates the agreement with the predicted result. At the onset of QPT the phase particle classically behaves, the quantum fluctuations vanish, i.e., the effective capacitance at $k=0$ diverges: the hopping amplitude between two successive minima of the potential—one even and the next odd—is zero. Conversely, the bandwidth remains finite, consistent with the analysis of \cite{korshunov}, since it originates from a small but non‑zero tunneling amplitude between two even (or two odd) minima, despite this hopping being extremely weak. 

\section{Regular contribution to the response function}
In this appendix, we outline the procedure used to extract the regular contribution to the correlation function, $\Psi_{Q, \mathrm{reg}}(\omega)$. We begin with the relation connecting $\Psi_{Q, \mathrm{reg}}(\omega)$ to the charge-charge correlation function
\begin{equation}\label{eqn:neweq}
    \frac{\hbar \omega}{2\pi }\int_0^{\infty} d\omega \, \Psi_{Q, \mathrm{reg}} (\omega)  \frac{\cosh\bigl(\omega(\beta \hbar -\tau)/2\bigr)}{\sinh \bigl( \beta \hbar \omega/2\bigr)} = \langle Q(\tau) Q(0)\rangle + \frac{D_{Q,M}- D_Q}{\beta},
\end{equation}
where $\tau > 0$ and $\tau < \hbar \beta$. Verification follows immediately from expanding all quantities in the Hamiltonian eigenbasis. 

The next step is to rewrite $\langle Q(\tau) Q(0)\rangle$ in terms of a correlation function of the phase and current operators: $\langle Q(\tau) Q(0)\rangle = C\langle \phi(\tau) I(0) \rangle$. This relationship is derived using $Q(\tau)=\frac{i C}{\hbar} [H,\phi](\tau)$, $H(\tau)=H$, the cyclic property of the trace, and $I=\frac{[H,Q]}{i \hbar}$. Finally, by integrating twice both members of Eq.~\ref{eqn:neweq} with respect to $\tau$ and using the sum rule for the regular part, $\Psi_{Q, \mathrm{reg}}(\omega)$, we obtain: 
\begin{equation}\begin{split}\label{eq:an_ric_eq_supp}
    &\frac{\hbar}{e^2} \frac{E_C^2}{\hbar^2} \frac{2}{\pi \omega}\int_0^{\infty} d\omega \, \frac{\cosh( \beta \hbar \omega/2) -  \cosh(\omega(\beta \hbar-2\tau)/2)}{\sinh(\beta \hbar\omega/2)}\Psi_{Q, \mathrm{reg}} (\omega)   = \langle (\varphi(\tau) - \varphi(0))^2 \rangle - \frac{D_Q}{\hbar^2 } \frac{E_C^2}{e^2} \frac{\tau}{\beta}(\hbar \beta - \tau),
\end{split}    
\end{equation}
This exact relationship allows us to express $\Psi_{Q, \mathrm{reg}}(\omega)$ in terms of a correlation function involving only phase operators, which can be easily determined via WLMC. Consequently, having determined $D_Q$ as described in the previous section, we invert Eq.~\eqref{eq:an_ric_eq_supp} using the MaxEnt~\cite{jarrell} implementation of Ref.~\cite{triqs_maxent} and as the default model a simple constant that satisfies the sum rule.

\end{document}